\documentclass[%
 aip,
 sd,%
 amsmath,amssymb,
 reprint,%
]{revtex4-1}

\usepackage{graphicx}
\usepackage{dcolumn}
\usepackage{bm}

\begin{document}

\preprint{AIP/123-QED}

\title{Semi-analytical approaches to study hot electrons in the shock ignition regime\footnote{Error!}}

\author{M. Afshari}
 \email{m.afshari@gsi.de}

\affiliation{ 
\begin{small}
GSI Helmholtzzentrum f{\"u}r Schwerionenforschung GmbH, Planckstra{\ss}e 1, 64291 Darmstadt, Germany
\end{small}
}%

\author{L. Antonelli}%
\affiliation{ 
Department of Physics, University of York, YO10 5DD York, England
}%

\author{F. Barbato}

\author{G. Folpini}%
\affiliation{Universite Bordeaux, CNRS, CEA, CELIA, UMR 5107, F-33405 Talence, France}

\author{K. Jakubowska}%
\affiliation{ 
Institute of Plasma Physics and Laser Microfusion, IPPLM, 01-497 Warsaw, Poland}%

\author{E. Krousky}%
\affiliation{ 
Institute of Plasma Physics CAS, v.v.i., 1782/3 182 00 Prague, Czech Republic}%

\author{O. Renner}%

\author{M. Smid}%
\affiliation{ 
Institute of Physics of ASCR, v.v.i., 1999/2 182 21 Prague, Czech Republic
}%

\author{D. Batani}%
\affiliation{Universite Bordeaux, CNRS, CEA, CELIA, UMR 5107, F-33405 Talence, France}

\date{\today}

\begin{abstract}
Hot electrons role in shock generation and energy deposition to hot dense core is crucial for the shock ignition scheme implying the need for their characterization at laser intensities of interest for shock ignition. In this paper we analyze the experimental results obtained at the PALS laboratory and provide an estimation of hot electrons temperature and conversion efficiency using a semi analytical approach, including Harrach-Kidder's model.
\end{abstract}

\maketitle

\section{Introduction}
\label{sec.1}
By increasing the gain and mitigating the constraints on the target fabrication and irradiation uniformity, the Shock Ignition (SI) approach to inertial confinement fusion  \cite{Bet07} should allow achieving nuclear ignition with current laser facilities \cite{Bat14a}$  ^{,}$ \cite{Per09}. In spite of such advantages, there are still many unresolved issues concerning the physics of shock ignition. One of the most important topics is the effect of parametric instabilities. They not only reduce the coupling of laser energy to the plasma (reflection of incident laser light), resulting in lower shock pressure, but  can also accelerate a fraction of electrons to high energies creating Hot Electrons (HE) \cite{Rib09}$  ^{,}$\cite{Koe13}. \\
HE have equivocal effect in SI:  if their energies are higher than $ \approx $ 100 keV, they can preheat the fuel making its compression more difficult and finally can prevent ignition. On the other hand, if they have energies below $ \approx $100 keV, they are unable to penetrate to the denser part of the target, and they may turn to be a positive factor by increasing laser-target coupling and shock pressure \cite{Bat12}. 
Laser plasma interaction produces HE by different physical mechanisms such as SRS (Stimulated Raman Scattering), TPD (Two Plasmon decay) and resonant absorption. All of them produce non mono-energetic electrons characterized by a distribution in energy f(E) and hence a hot electron temperature, T$ _{HE} $. The energy conversion and HE temperature are critical parameters in order to evaluate the positive vs. negative effects of HE in SI.\\
In order to study the physics of HE generation in an intensity regime relevant to SI, we performed an experiment at the Prague Asterix Laser System laboratory, PALS in Prague Czech republic. One of the goals of the experiment was the study of generation and propagation of HE and their role in shock wave generation \cite{Bat12}. The objective of this paper is the analysis of the experimental results with semi-analytical approaches, in particular Harrach-Kidder's model \cite{Har81}. This model is particularly adapted to describe the propagation of HE in the target when HE energy is not high and therefore the motion of electrons become quickly isotropic due to collisions in the target. Although Monte-Carlo simulations are a powerful tool for analyzing data on HE dynamics, nevertheless the use of semi-analytical models may represent a quick alternative for estimating the parameters of HE distribution, with the advantage of suggesting information on the physical mechanisms at play.  \\
The paper is organized as follow: Sec.2 contains the set-up of the experiment. The analysis of raw X-ray images to determine the K$  \alpha$ spot size and the total number of emitted K$  \alpha$ photons are illustrated in Sec.3. In Sec.4 we present the experimental results and compare  the characteristic of HE emitted at 1$ \omega $/3$ \omega $ laser frequencies. In Sec.5 we estimate the HE energy first by calculating the range and relating it to HE energy using the ESTAR database \citep{EST05}  and then with the semi-analytical model developed by Harrach \& Kidder. Sec.6 contains the conclusions.\

\section{Experiment set-up and diagnostic}
\label{sec. Experiment set-up and diagnostic}
\begin{figure}[!htp]               
\centering
\includegraphics[trim={2.5cm 3cm 1cm 2.8cm},clip,scale=.6]{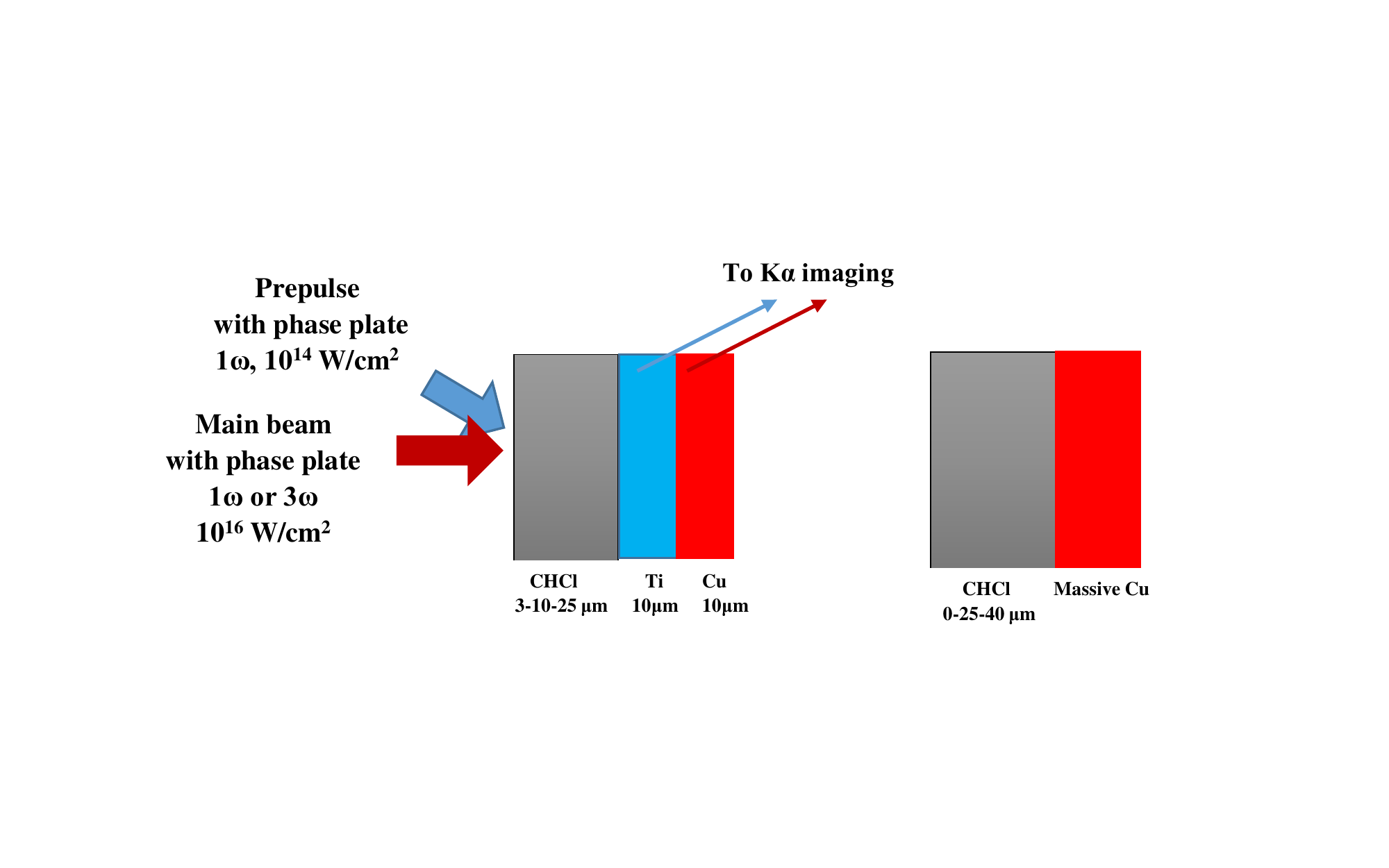}
\caption{\small  Schematic view of the thin multilayer targets (left) and thick Cu targets (right).}
 \label{target}
 \end{figure}

In this Section we describe the set-up and the diagnostics. In the experiment we used two laser beams. The first one is the PALS main laser beam operating either at 1$ \omega $ or 3$ \omega $ harmonics ($ \lambda $= 1315 nm/438 nm, maximum energy 440/170J, pulse duration 300 ps) focused onto the target with  100$\mu$m focal spot. The intensity of the main laser pulse was varied by changing the laser pulse energy, whereas pulse duration and focal spot size were kept fixed throughout the experiment. The maximum available intensity was 2$\times$10$ ^{16} $ W/cm$ ^{2} $ at 1$ \omega $ and 9$\times$10$ ^{15} $  W/cm$ ^{2} $ at 3$ \omega $. The second beam, used on some shots, was an auxiliary beam ($ \lambda $=1315 nm at 1$ \omega $, energy up to 100 J, pulse duration 300 ps, I $ \approx $10$ ^{14} $ W/cm$ ^{2} $) focused onto the target with the Gaussian focal spot ($ \approx $ 300 $\mu$m) incident at 30$ ^\circ $ from target normal in the vertical direction. For the auxiliary beam we chose 1$ \omega $ irradiation because in this case the laser beam generates a plasma with higher temperature at the critical surface in comparison to plasma produced  at 3$ \omega $. In this way we tried to compensate the low energy available in  the beam. The main pulse irradiated the target after the auxiliary beam with delays Δt = 0, 300, 600 and 1200 ps. In fact the auxiliary beam creates a long-scale plasma (preplasma) and then the main laser pulse at 1$ \omega $ or 3$ \omega $ focused inside the preplasma produces the HE beam and launches a shock wave.\\
To ensure a better focal spot uniformity, phase plates were used on both beams. On a few shots at 3$ \omega $, the phase plate on the main beam was removed to reach higher intensities (up to 3$\times$10$ ^{16} $ W/cm$ ^{2} $) in a focal spot with the average diameter $ \approx $ 60 $\mu$m. \\
During the experiment we used different targets but two types of targets were dedicated specifically to HE studies. The first type of targets were thin multilayer targets: CHCl-Ti-Cu. The first layer was a Chlorinated plastic (Parylene-C, C$ _8 $H$ _7 $Cl) with different thicknesses (3-10-25 $\mu$m). The CHCl layer was followed by two thin  layers of high Z material (Ti, Cu) used as traces layers for the fluorescence emission resulting from the interaction of HE. The second type of targets were thick Cu targets  either bare or coated with a CHCl layer on the laser side with thicknessess 25 and 40 $\mu$m. The thick Cu layers exclude the possibility of HE refluxing. Fig.\ref{target} shows a schematic of targets.\\
\\
To characterize the HE generation in the plasma we used spherically bent crystals to image the K$\alpha$ spot onto the X-ray films \cite{Balu14}. 
\begin{figure}[!htp]   
\centering
\includegraphics[trim={.5cm 15cm 0cm 4cm},clip ,scale=.35]{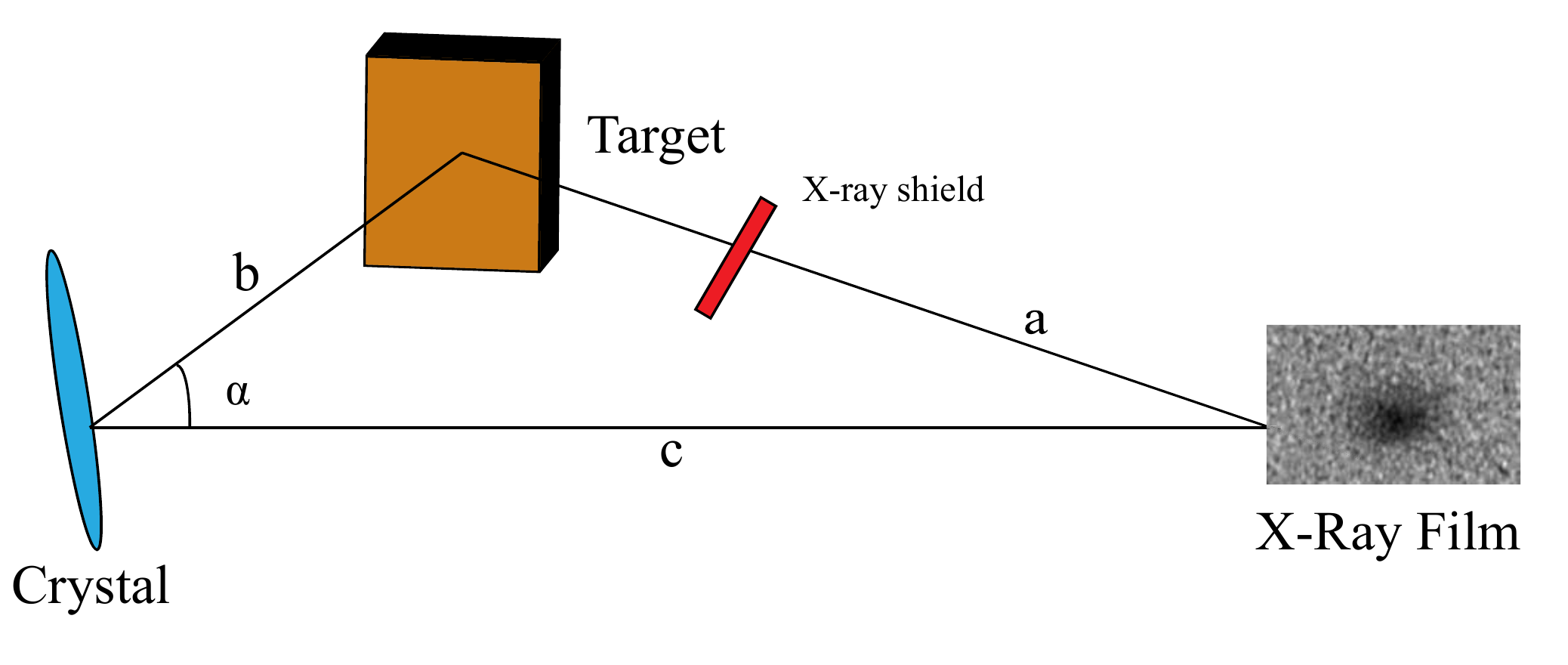}
\includegraphics[trim={10cm 0cm 0cm 0cm},clip ,scale=.3]{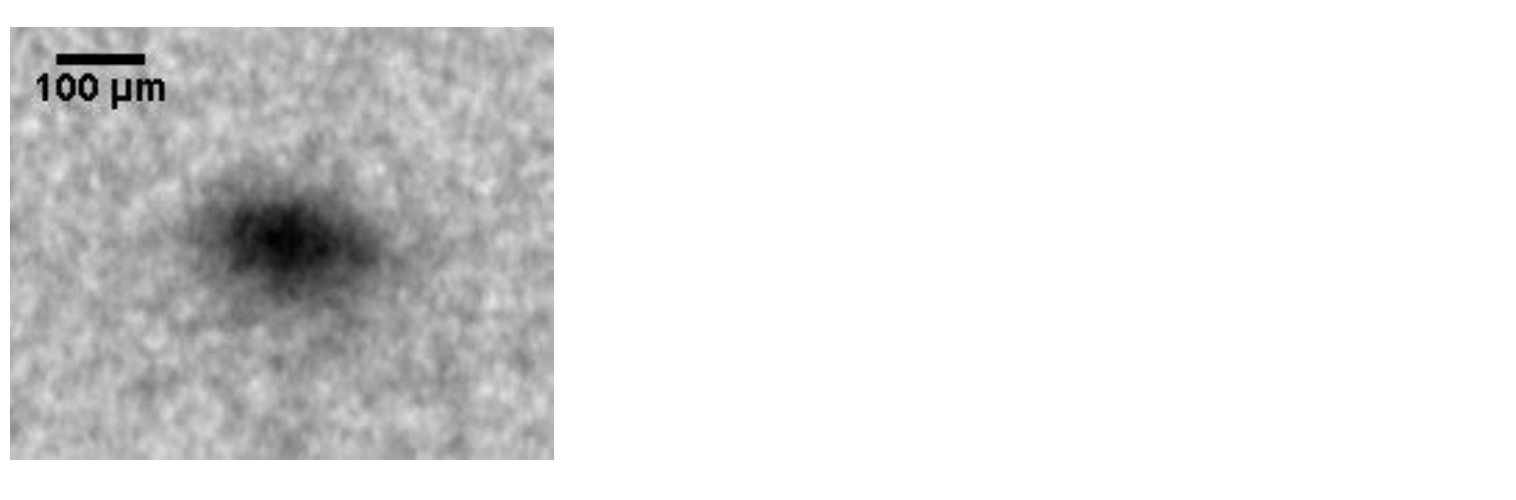}
\caption{\small Left: Set up for K$\alpha$ imaging measurement. The X-ray shield protects the X-ray film from direct irradiation of the target. The distances of Target-Film (a), Target-Crystal (b) and Crystal-Film (c) are 22 cm, 30 cm and 51.9 cm respectively. The magnification of the system is M =1.73. \newline
Right: Typical Cu K$  \alpha$ spot (black spot) as recorded on the X-ray film. }
\label{ka-image}
 \end{figure}
Fig.\ref{ka-image} shows the experiment setup and a typical image of the X-ray film. The films used in our experiment (Kodak Industrex $  \times$ AA400 film) are calibrated \cite{Rid89}. Images (Fig.\ref{ka-image}, right) show the geometry of the K$  \alpha$ spot and hence the geometry of the HE beam crossing the Ti/Cu layers. They also allow to extract the total number of K$  \alpha$ photons which irradiated the film. To estimate the total number of K$  \alpha$ photons emitted from the target on 4$  \pi$, we multiply the total number of K$  \alpha$ photons detected on the X-ray film by the transfer characteristics of the imaging system, $ \zeta $

\begin{equation}
\zeta=\frac{4\pi}{R \:T \: \Omega}   
\end{equation}
where 
\begin{description}

\item[$\bullet$]R is Integrated reflectivity of the crystal, which for Cu is R$ _{Cu} $=47.02 $\mu$rad at wavelength 1.5406 \AA \ and for Ti is R$ _{Ti} $= 906.0 $\mu$rad at wavelength 2.748 \AA.  
\item[$\bullet$]T is the transition of the filters placed before the X-ray film, 60 $ \mu m $ Al+ 20 $ \mu m $ mylar. The overall transmission of K$ \alpha $ through these filters for Cu is  T$ _{Cu} $=0.45 and for Ti is T$ _{Ti} $= 0.014.
\item[$\bullet$] $ \Omega$ is the collection solid angle given by $ \Omega$= \begin{Large}
$\frac{S sin(\theta_{\beta})}{b^{2}}$
\end{Large} where S is the area of the crystal surface (the radius of the crystal surface is 1.2 cm), $  \theta_{\beta}$ is the Bragg
angle of the crystal (88.15$^{\circ}$) and b is the target-crystal distance (30cm).
\end{description}
Taking into consideration all above parameters, the transfer characteristics of the imaging system is estimated to be  $ \zeta_{Cu} $=$ 5.2 \times 10 ^{5}  $ for Cu and $ \zeta_{Ti} $=$ 1.6 \times 10 ^{6}  $ for Ti. For instance, this implies that one Cu K$ \alpha $ photon collected on the X-Ray film corresponds to $ 5.2 \times 10 ^{5}  $  photons emitted from the target in 4$  \pi$.\\
We have to mention the coronal x-ray emissions, which are the bremsstrahlung emissions of electrons from the tail of electron distribution in the corona with energy above the K$ \alpha $ emission, should be very negligible in comparison to the x-ray emissions due to the interaction of HEs with the target. The Cu K$ \alpha $ pumping by coronal x-rays has been analyzed in previous studies\citep{ren16} where for analogous experimental conditions it was demonstrated that even at uncoated Cu targets, where the coronal temperature is rather high ($ \leq $2 keV, as determined experimentally and supported by Multi-2D hydrodynamic modeling \citep{smid14}), the K-shell excitation due to the corona radiation can be neglected, being at the level of 5\% compared to the effect of HEs with the characteristic temperature of 30 keV. At the same time, the coronal temperature at plastic coated targets is much lower, typically at the level of 700 eV \citep{pis15}, i.e. the effect of K-shell pumping due to the coronal emission is still smaller. We note that the contribution of HE induced continuum is much weaker in comparison to the action of coronal radiation, thus the K-shell excitation is governed by HEs and the effects of radiation originating from the laser matter interaction can be neglected.

\section{Experimental Results}
We studied the variation of the K$\alpha$ spot size and the K$ \alpha$ photon numbers as a function of:
\begin{description}

\item[$\bullet$] Energy of main laser beam;
\item[$\bullet$]Preplasma (energy of auxiliary laser beam and delay between the two beams);
\item[$\bullet$]Thickness of the plastic layer before the tracers (Cu and Ti); 
\end{description}

\subsection{ Measurements of K$\alpha$ Spot size vs. Laser Energy} 
\begin{figure}[!htp]  
\centering
\includegraphics[trim={0cm 7cm 0cm 7cm},clip ,scale=.45]{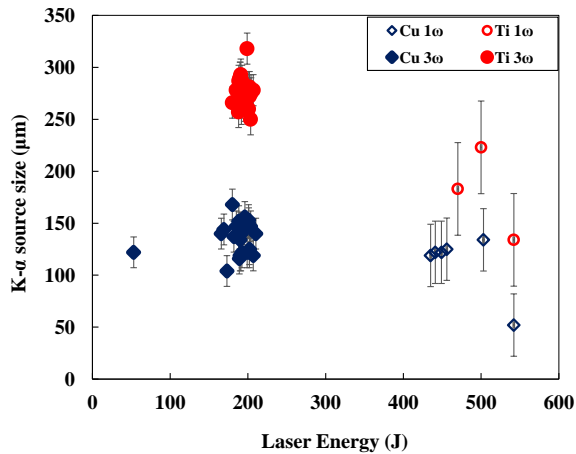}
\caption{\small K$\alpha$ spot size vs. main laser energy for thin multilayer targets.}
 \label{ka_vs_laser_target1}
 \end{figure}

To determine the effect of the laser energy on K$\alpha$ spot size, we plot the K$\alpha$ spot sizes of thin targets vs. the laser energy for both 1$ \omega $/3$ \omega $ frequencies (Fig.\ref{ka_vs_laser_target1}).
 \\
 On average the size of the Cu K$\alpha$ spot is $\approx 130 \mu $m which is about 1.3 times the size of the focal spot. As we can see the Cu K$\alpha$ spot size is comparable at both 1$\omega$/3$\omega$ frequencies. The Ti K$\alpha$ spot size is bigger than the Cu one at both 1$\omega$/3$\omega$ frequencies, typically a factor 1.5-3. This might at first seems strange because one can expect the size to increase with depth, due to the divergence of HE in the target. Nevertheless it has to be considered that the Cu is the third layer and that Cu K$\alpha$ emission (8027 eV) is more energetic than Ti K$\alpha$ emission (4504 eV). Hence only the more energetic electrons can reach the Cu layer and excite it and these more energetic electrons are probably emitted with lower angular divergence.\\

\subsection{Measurement of  K$\alpha$ photons vs. Laser Energy} 
\label{sec.3.2} 
\begin{figure}[!h]    
\centering
   \includegraphics[trim={0cm 8.5cm 0cm 7cm},clip ,scale=.45]{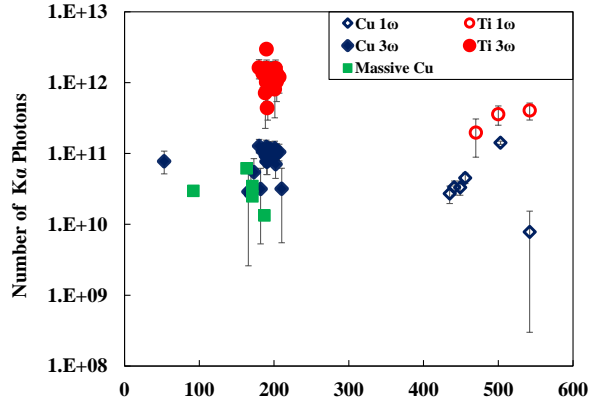}
\caption{\small Number of K$  \alpha$ photons versus main laser energy for thin targets at both 1$\omega$/3$\omega$ frequencies and thick Cu targets at 3$\omega$ frequency.}
\label{NPhs-vs-laser}
 \end{figure}

The variation of the number of Cu and Ti  K$  \alpha$ photons emitted over the whole solid angle, 4$ \pi $,  for thin targets is plotted versus laser energy for both 1$ \omega $/3$ \omega $ frequencies in Fig. \ref{NPhs-vs-laser}.
 \\
On average the number of  K$  \alpha$ photons emitted over 4$ \pi $ are: (1$\pm$ 0.5)$ \times 10 ^{12}$ for Ti at $3\omega $, (4 $  \pm$ 1) $ \times 10 ^{11}$ for Ti at $1\omega $, (1 $\pm$ 0.2)$ \times 10 ^{11}$ for Cu at $3\omega $, and (5 $  \pm$ 4) $ \times 10 ^{10}$ for Cu at $1\omega $. \\
 As expected the number of Ti  K$  \alpha$ photons is higher than the number of Cu  K$  \alpha$ photons at both 1$\omega$ and 3$\omega$. Indeed Ti is the second layer (therefore more HE arrive there in comparison to the Cu layer) and also Ti K$  \alpha$ is excited at lower energies. \\
Fig.\ref{NPhs-vs-laser} also shows the variation of the number of emitted Cu  K$  \alpha$ photons for thick Cu targets at 3$  \omega$. The smallest number of Cu K$\alpha$ photons is obtained with the thick Cu targets with plastic over layer. This seems to point out to some effects of HE refluxing in the thin targets (effect which is of course not present in the thick targets).

\subsection{Study of the Effect of Preplasma on K$\alpha$ Spot size and K$\alpha$ intensity} 
\begin{figure}[!htp]    
\centering
\includegraphics[trim={0cm 7cm 0cm 7.5cm},clip ,scale=.45]{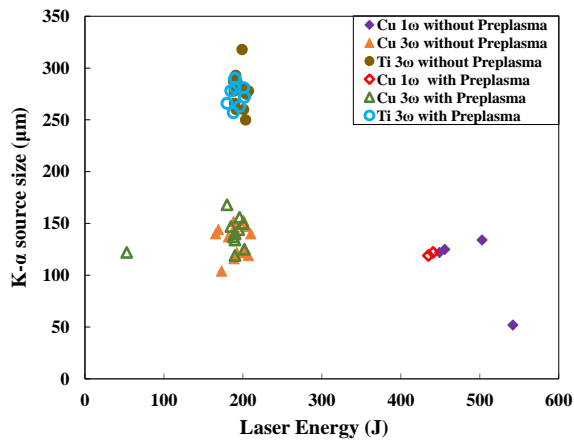}
\caption{\small  Effect of preplasma on K$\alpha$ spot size for thin targets at both 1$\omega$/3$\omega$ frequencies.}
\label{ka-preplasma}
\end{figure}

\begin{figure}[!htp]    
\centering
\includegraphics[trim={0cm 8cm 0cm 7.5cm},clip ,scale=.47]{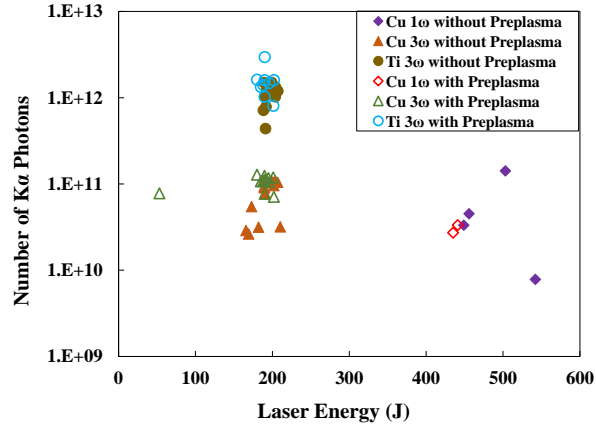}
\caption{\small Effect of preplasma on the number of K$  \alpha$ photons for thin targets at both 1$\omega$/3$\omega$ frequencies.
 }
 \label{NPhs-preplasma}
\end{figure}

To show the effect of preplasma on K$\alpha$ spot size we divided the data into two groups, with and without preplasma. We have plotted in Fig.\ref{ka-preplasma} the variation of the K$\alpha$ spot size for these two groups versus laser energy at both 1$\omega$/3$\omega$ for thin targets. 
\newline
As we can see all K$\alpha$ spots with or without preplasma have approximately the same size, showing that the preplasma has no appreciable effect on the K$\alpha$ spot size at both 1$\omega$/3$\omega$ frequencies.\\
To check the effect of preplasma on the number of  K$  \alpha$ photons
we plotted in Fig.\ref{NPhs-preplasma} the variation of  K$  \alpha$ photon numbers vs. the laser energy at both 1$\omega$/3$\omega$ for  thin targets with and without preplasma. 
 \\
Again, the preplasma seems to have no main effect on K$  \alpha$  emission. 
 
\subsection{Study of the effect of plastic thickness on K$\alpha$ emissions} 
\begin{figure}[!htp]    
\centering
\includegraphics[trim={1cm 0cm 0cm 0cm},clip ,scale=.33]{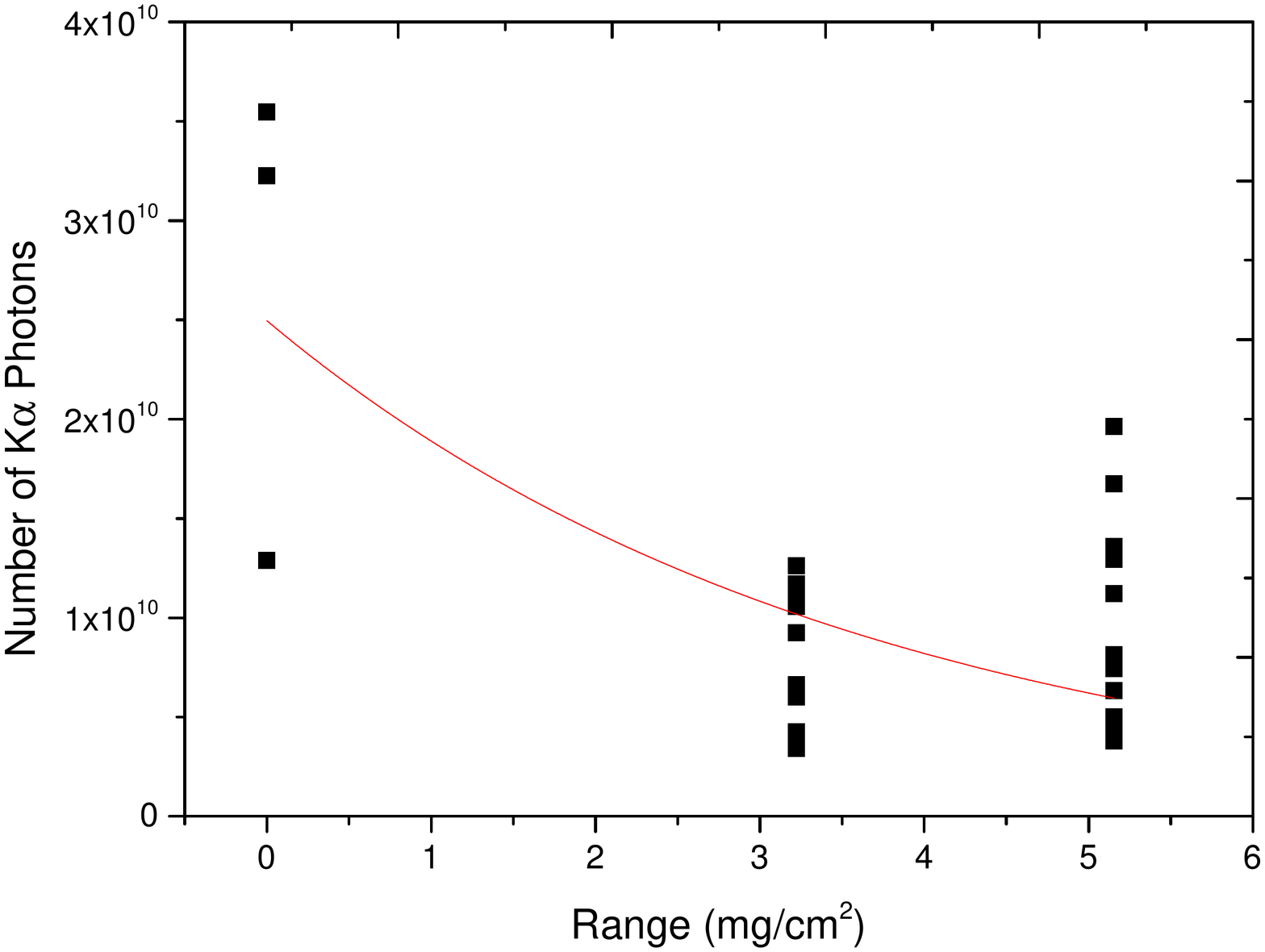}
\caption{\small Number of K$  \alpha$ photons vs. range of CHCl for thick Cu targets at 3$  \omega$. Data points are interpolated with an exponential function.}
\label{Spreading-angel-massive}
\end{figure}

\begin{figure}[!htp]    
\centering
\includegraphics[trim={0cm 0cm 0cm 0cm},clip ,scale=.3]{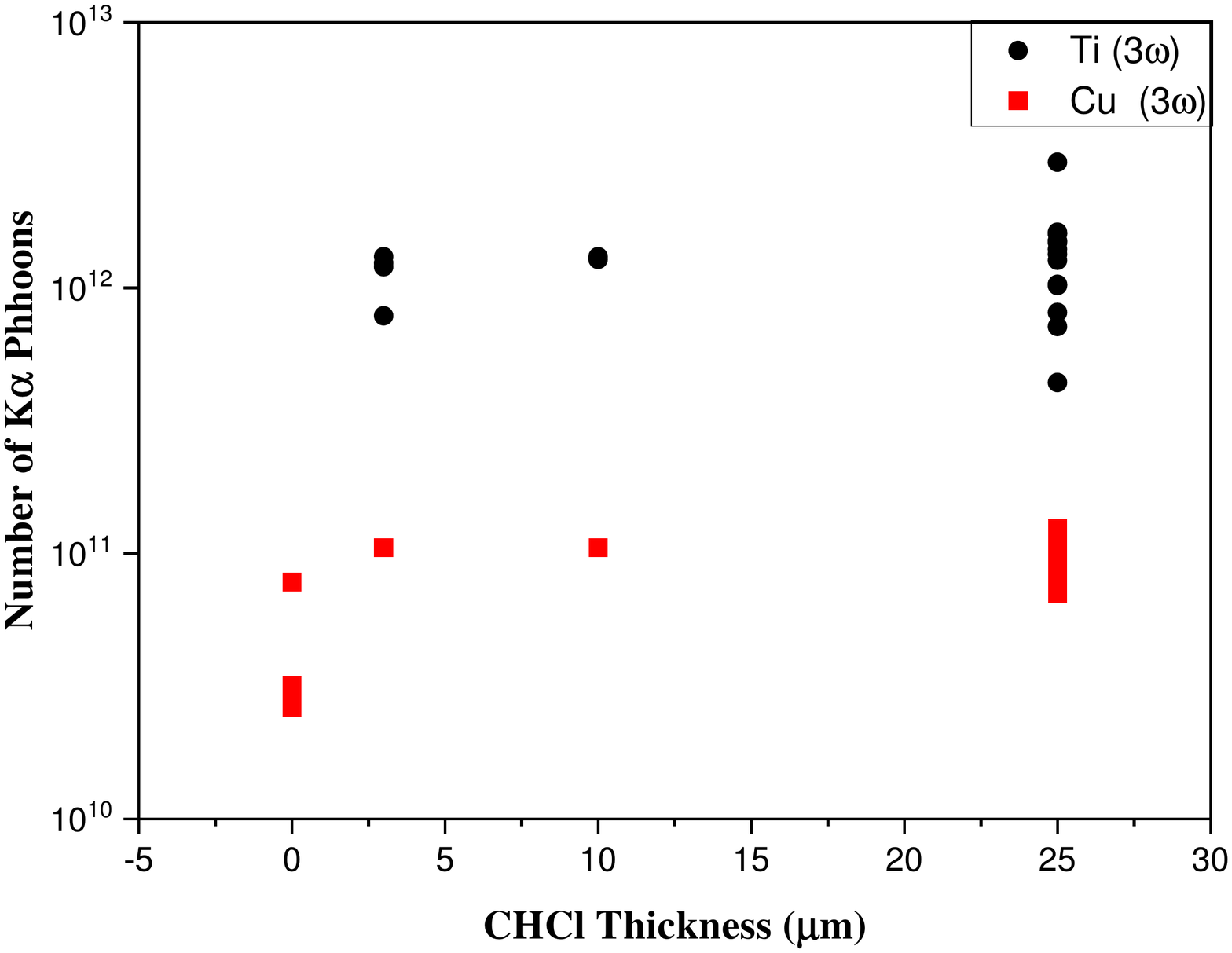}
\caption{\small Number of K$  \alpha$ photons vs. CHCl thickness for thin targets at 3$\omega$.}
\label{NPhs-vs-CHCl-thickness}
\end{figure}

In this section we show the variation of the number of  K$\alpha$ photons vs. the thickness of the CHCl overlayer. Fig.\ref{Spreading-angel-massive} shows the case of the thick Cu targets.  
 \\
With increasing the CHCl thickness the K$\alpha$ signal is decreasing. From such decrease we can evaluate the average energy of HE as we will do in the next section.\\
We repeated the analysis for thin multilayer targets, Fig.\ref{NPhs-vs-CHCl-thickness}.
 \\
While Fig.\ref{Spreading-angel-massive} shows a clear decrease, fig.\ref{NPhs-vs-CHCl-thickness} shows a quasi flat behaviour. The two might not necessary be in contradiction because Fig.\ref{Spreading-angel-massive} extends to larger CHCl thickness so it allows to better see any decrease. In addition in Fig.\ref{NPhs-vs-CHCl-thickness}, the decreasing slope could be probably hidden inside the large  shot-to-shot fluctuations and the low statistics. \\
With respect to Fig.\ref{NPhs-vs-CHCl-thickness} we also notice that the number of Cu K$  \alpha$ photons emitted from pure Cu targets (no overlayer) is less than that obtained when the plastic overlayer is present. This is indeed expected when the laser beam directly interact with Cu, since it produces a hot ablated plasma where Cu is strongly ionized and emit thermal X-ray or shifted K$\alpha$ lines instead of cold K$\alpha$ line which is collected by crystal.

\section{Estimation of hot electron average energy} 
\label{sec.4}
\subsection{Exponential interpolation}
\label{sec.ESTAR database}
\begin{figure}[!htp]    
\centering
\includegraphics[trim={0cm 6cm 1cm 6cm},clip ,scale=.4]{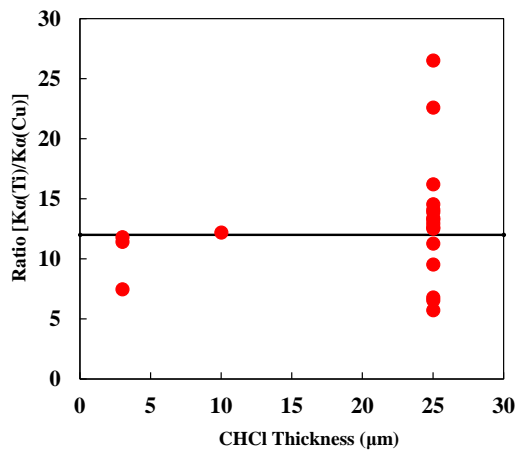}
\caption{\small  Ratio of the number of K$ \alpha $ photons of Ti to the number of K$ \alpha $ photons of Cu vs. CHCl thickness for thin targets at 3$\omega$. The solid line represents the average ratio.}
\label{NPhs_ratio-spreading_angel}
\end{figure}

The data of Fig.\ref{Spreading-angel-massive} can be interpolated with an exponential function, 
\begin{equation}
N_{e}(x)= N_{0} \: exp (-x/R)
\label{ESTAR FIT}
\end{equation}
where N$ _{0} $  is the number of photons which would ideally be emitted from a target without overlayer and x the areal density before the tracer layer \cite{Mor10}. By assuming that the number of emitted K$ \alpha $ photons is roughly proportional to the number of HE reaching the tracer layer, we can estimate the range (g/cm$ ^{2} $), R, of HE. This is a crucial parameter since shows how much HE can penetrate into the target and hence allows to estimate their average energy. For this purpose we can use the ESTAR database which relates the initial energy of HE to range\cite{EST05}. The simplest way to evaluate HE energy is assuming a mono energetic electron beam instead of considering a realistic electron energy distribution (i.e. all of HE have the same energy corresponding to the average energy in the distribution) and assuming that all electrons travel in straight lines (instead of realistic trajectories characterized by scattering and angular divergence).\\
For thick targets, Fig.\ref{Spreading-angel-massive}, we fitted the curve with an exponential function, Eq. 2, which gives the initial number of photons N$ _{0} $= (2.5 $ \pm $ 0.3)$ \times 10^{10} $ and a range R= 3.6$\pm  $0.6 (mg/cm$ ^{2} $).According to the ESTAR database such range corresponds to HE with energy $ T_{HE} \approx$ 41.5$ \pm 6 $ keV.\\ 
For thin targets, the number of  K$  \alpha$ photons doesn't seem to change appreciably with increasing the plastic thickness from 3 to 25 $  \mu$m. Hence, we can't evaluate energy of HE with the same method. For these targets we use a different approach based on the ratio of the number of K$ \alpha $ photons of Ti to the number of K$ \alpha $ photons of Cu vs. CHCl thickness. Fig.\ref{NPhs_ratio-spreading_angel} shows that, taking into account the very large shot-to-shot fluctuations, the ratio (Ti K$ \alpha $/Cu K$ \alpha $) can be assumed to be constant and $ \approx 12 \pm 7 $. 
\newline
In the following, we assume the ratio of K$ \alpha $ photons is equal to the ratio of the number of electrons which arrive to the Ti layer to those arrive to the Cu layer. The number of electrons reaching the Ti layer is

\begin{equation}
N  _{Ti,0} = N  _{0} \ exp (- \frac{l_{CH}}{R_{CH}(E _{0})})
\label{Eq.Number-HE-CHCl} 
\end{equation} \newline
where $ N  _{0} $ is the initial number of electrons, l$ _{CH} $ is the CHCl areal density and R$ _{CH} $(E$ _{0} $) is the range in CHCl which is a function of energy and E$ _{0} $  is the initial energy of HE (again assumed mono energetic). The ESTAR database also gives the energy loss, $  \Delta$E$ _{CHCl} $, in the CHCl layer. Accordingly, electrons with initial energy E$ _{0} $ arrive to the Ti layer with energy E$ _{Ti} $= E$ _{0} $ - $  \Delta$E$ _{CHCl} $. We assume on average the Ti K$ \alpha $ photons are emitted at half of the Ti layer

\begin{equation}
N  _{Ti} = N  _{Ti,0} \  exp (-\frac{l_{Ti}/2}{R_{Ti}(E_{Ti})})
\label{Eq. Number HE Ti} 
\end{equation}\newline
where l$  _{Ti}$ is the  areal density of the Ti layer and R$  _{Ti}$(E$ _{Ti} $) is the range in Ti for electrons with energy E$ _{Ti} $.\\
The fraction of more energetic electrons of $ N  _{Ti,0} $  which pass Ti layer and reach the Cu layer is given by 

\begin{equation}
N  _{Cu,0} = N  _{Ti,0} exp (- \frac{l_{Ti}}{R_{Ti}(E_{Ti})} )
\end{equation} \newline
They reach the Cu layer with an energy E$ _{Cu} $= E$ _{Ti} -   \Delta$E$ _{Ti} $ where $ \Delta$E$ _{Ti} $ is the energy loss in Ti layer. For the Cu layers we assume that on average Cu K$ \alpha $ photons are emitted at the depth equal to $ R_{Cu} $/2, where R$  _{Cu}$(E$ _{Cu} $) is the range  in Cu for electrons with energy E$ _{Cu} $. Hence we have 

\begin{equation}
N  _{Cu} = N  _{Cu,0} \  exp (-\frac{R_{Cu}(E_{Cu})/2}{R_{Cu}(E_{Cu})}) \approx 0.6 \times N  _{Cu,0} 
\label{Eq. Number HE Cu} 
\end{equation}\newline
accordingly the ratio equals to

\begin{equation}
\begin{Large} 
  \frac{N  _{Ti} }{N  _{Cu}} = 1.7 \times  exp (\frac{l_{Ti}/2}{R_{Ti}(E_{Ti})})
 \end{Large}
\label{Eq.analytic-ratio2} 
\end{equation}\newline
which depends on the initial energy of HE(E$ _{0} $), the areal density of the CHCl layer (E$ _{Ti} $= E$ _{0} -\Delta E _{CHCl} $) and the areal density of the Ti layer (E$ _{Cu} $= E$ _{0} -\Delta E _{CHCl} -\Delta E _{Ti}  $). \\
Also the number of emitted K$  \alpha$ photons is proportional to 

\begin{equation}
 N(K  \alpha) \approx  N \: \sigma(E) \: n \: \omega_{K} \: l
\label{Eq.Nph by HE}
\end{equation}
\newline
where N is the number of HE with energy sufficient to generate K$ \alpha $ emissions, $ l$(cm)  is the thickness of each layer, $ \omega_{K}  $ is the florescence yield of K shell which for Cu  is $ \omega_{(K, Cu)}= 0.38  $ and for Ti is $ \omega_{(K, Ti)}= 0.17  $, n($ cm^{-3}$) is the ion density which for Cu is n$ _{Cu} $ = $8.4\times 10^{22}cm ^{-3} $ and for Ti is n$ _{Ti} $ = $5.7\times 10^{22}cm ^{-3} $. The K$ \alpha $ cross section, $ \sigma (cm^{2} ) $, is energy dependent and given by \cite{landa13} 
\begin{equation}
   \sigma (E)(cm^{2} )= 7.92 \times 10^{-14} \frac{1}{EE_{K}} ln(\frac{E}{E_{K}})   
\label{Eq.k-alpha corss section}
\end{equation}\\
where E is the energy of HE (in eV) and $ E_{K} $ is the K$ \alpha $ ionization energy which for Cu is about 8987 eV. Since the CHCl thickness varies and $  \sigma $ is energy dependent we calculate its values depending on the HE energy. In general, we find for Cu, $ \sigma_{Cu} \approx (2-4)\times10^{-22}$  $ cm^{2} $ and for Ti,  $ \sigma_{Ti} \approx (8-9)\times10^{-22}$  $ cm^{2}$ .\\ 
Finally the ratio is proportional to
\begin{equation}
\begin{Large} 
 \frac{N(K  \alpha)_{Ti}}{N(K  \alpha)_{Cu}} \approx
  \frac{N  _{Ti} }{N  _{Cu}}  \times   \frac{\sigma_{Ti} \ n_{Ti} \ \omega_{k,Ti} \ l_{Ti}}{\sigma_{Cu} \ n_{Cu} \ \omega_{k,Cu} \ L_{Cu}}
 \end{Large}
\label{Eq.analytic-ratio2} 
\end{equation}
\newline
where $ L_{Cu} $ is the penetration depth in Cu layer. Table.\ref{table.ratio} summarises the results.
\begin{table*}[!htp]
\centering
\caption{\small  Ratio of the number of K$ \alpha $ photons of Ti to the number of K$ \alpha $ photons of Cu following Eq.\ref{Eq.analytic-ratio2}. l$ _{CHCl} $= 10$ \mu m$ is chosen for this analysis.}
\begin{tabular}{  >{\centering}m{.75cm}  >{\centering}m{1.3cm}  >{\centering}m{.75cm}    >{\centering}m{.75cm}  >{\centering}m{1.3cm} >{\centering}m{.75cm}  >{\centering}m{1.cm}>{\centering}m{.75cm} >{\centering}m{.75cm} >{\centering}m{1.cm} >{\centering}m{.75cm} >{\centering}m{1.3cm} >{\centering}m{1.cm} >{\centering}m{.75cm}>{\centering}m{.75cm}  } 
\hline 
\hline 
  E (keV) & R$ _{CH} $ \\ (mg cm$ ^{-2} $) & $  \Delta$E (keV) & E$ _{Ti} $ (keV) & R$_{Ti}$  \\ (mg cm$ ^{-2} $) & 
 N$ _{Ti,0} $ & $\sigma_{Ti}$ ($ 10^{-22} $) (cm$ ^{2} $) & N$ _{Ti} $ & $  \Delta$E$_{Ti}$ (keV) &  E$ _{Cu} $ (keV)& N$ _{Cu,0} $ & R$ _{Cu} $  \\  (mg cm$ ^{-2} $) & $\sigma_{Cu}$ ($ 10^{-22} $) (cm$ ^{2} $) &  N$ _{Cu} $ & $ \frac{N \:_{Ti}}{N \:_{Cu}} $    \tabularnewline 
\hline
40 & 3.2 & 9.3 & 30.7 & 2.8 & 0.67 & 11.6 & 0.44 & 29.1 & 1.6 & 0.20 & 0   & 0       & 
0 & 0 \tabularnewline 
50 & 4.7 & 7.9 & 42.1 & 4.8 & 0.76 & 9.5 & 0.62 & 23.2 & 18.9 & 0.39 & 1.3 & 2.6 & 0.24 & 19.2 \tabularnewline
52 & 5.0 & 7.7 & 44.3 & 5.2 & 0.77 & 9.1 & 0.65 & 22.4 & 22   & 0.42 & 1.7 & 3.0 &   0.25 & 11.6 \tabularnewline
54 & 5.4 & 7.5 & 46.5  & 5.7 & 0.79 & 8.8 & 0.67 & 21.7 & 24.9 & 0.45 &  2.1 & 3.3 & 
0.27  & 8.0 
\tabularnewline
\hline
\hline
\end{tabular}
\normalsize   
\label{table.ratio}
\end{table*}\\
The analysis estimate HE energy about T$ _{HE} $ $ \approx $ 52$ ^{+9} _{-5} $ keV. Taking into account the error bars, this result is comparable with what obtained from Fig.\ref{Spreading-angel-massive}.

\subsection{Harrach-Kidder model} 
\label{sec.H-k model}
\begin{figure}[!htp]    
\centering
\includegraphics[trim={0cm 1cm 0cm 2cm},clip ,scale=.3]{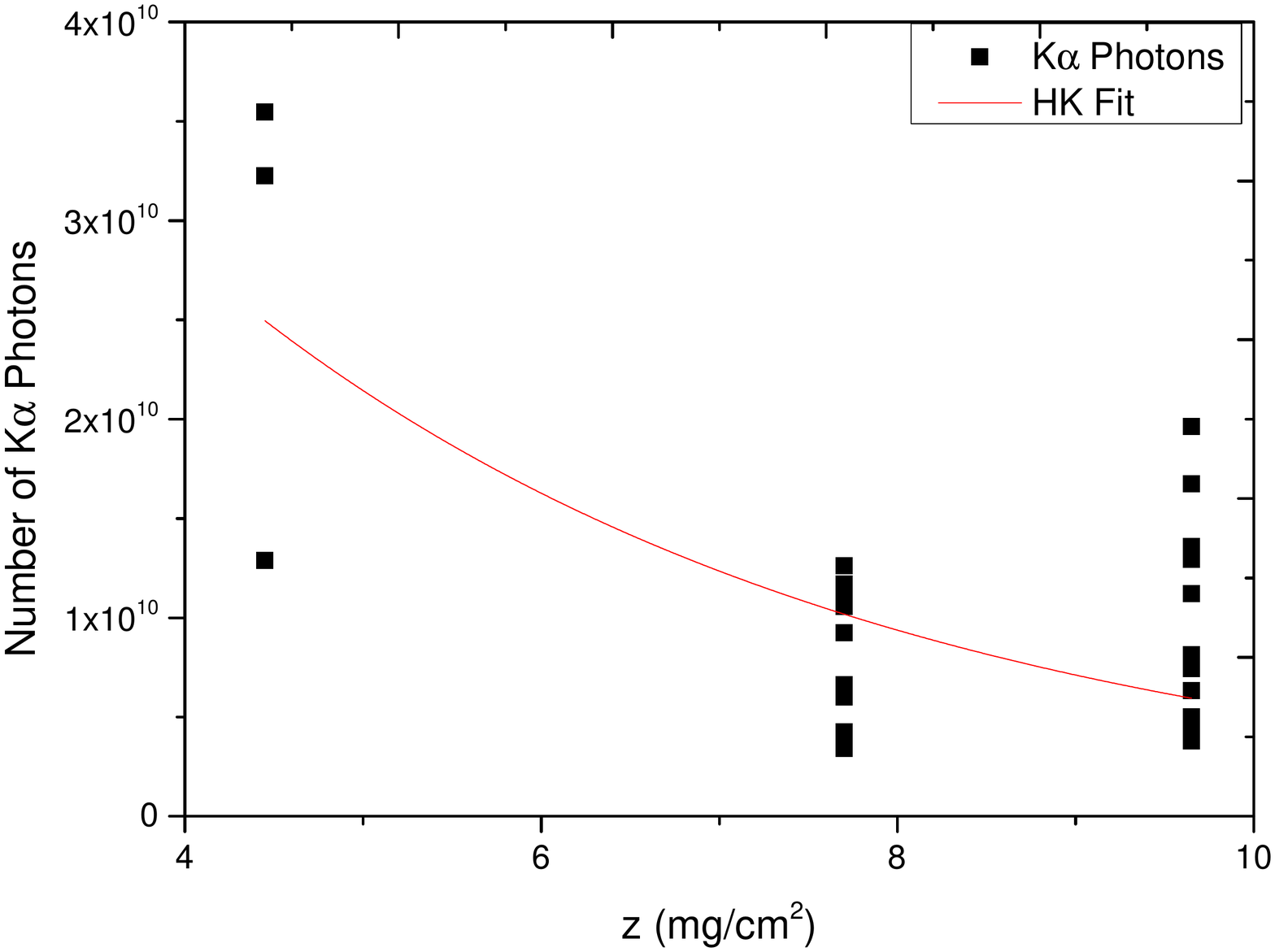}
\caption{\small  Number of K$ \alpha $ photons vs the total areal density  of target (Eq.\ref{Eq. target total length}) and its fit according to H-K model (Eq.\ref{Eq. H-K FIT}). }
\label{fig.HK-fit}
\end{figure}

In this section we estimate the HE energy using the semi-analytical model developed by Harrach \& Kidder \cite{Har81}.\\
The advantage of H-K model with respect to other analytical models \cite{Cap79}$ ^{,} $ \cite{Lee80}  is that these models don't apply to targets with low atomic number while H-K model not only benefits from simple analytical formula (which give better physical intuition) but also can be used for materials with any atomic numbers. More importantly its predictions are in good agreement with the experimental results for low HE energy \cite{Har81}.\\
In the model, the HE source is planar, isotropic and characterized by a Maxwellian velocity distribution with temperature T$ _{HE} $. A fraction of HEs penetrate directly towards the dense target and deposit their energy which is calculated using Spencer energy deposition function \cite{Spe59}. The energy deposited at distance z into the target is proportional to 
\begin{equation}
 N(z) \approx N  _{0}  exp [-  \beta\sqrt{z/R} ]
\label{Eq. H-K FIT}
\end{equation}  
where N$ _{0}$ is the initial number of electrons and $  \beta$ is a parameter which characterizes the propagating material, R is the range and z is the distance travelled inside the target in normal direction to the surface. Both R and z are measured in (g/cm$ ^{2} $) unit. The relation between the range and HE temperature, T$ _{HE} $, is
\begin{equation}
R=b (T _{HE})  ^{1+\mu} 
\label{Eq. H-K temperature} 
\end{equation} 
where T$ _{HE} $ in (keV) and parameters b and $ \mu $ depend on the propagating material. To apply Eqs.\ref{Eq. H-K FIT} and \ref{Eq. H-K temperature}, we need to know the values of $ \beta $, b, and $ \mu $. We choose the values of carbon: $  \beta_{CHCl}$=1.85, b$ _{CHCl} $=4.6$  \times$10$ ^{-6} $, and $  \mu_{CHCl}$=0.78 \cite{Har81}. This choice is justified by the fact that in the plastic layer, in front of the target, where HE generated and propagated, mostly carbon atoms are responsible for the HE stopping power \cite{Bat00}.\\
In the following we consider only thick Cu targets. In order to apply H-K model we should convert the thickness of the thick targets (CH+Cu) to the equivalent total areal density (g/cm$ ^{2} $) as needed in Eq.\ref{Eq. H-K FIT}. To do that we multiply the length of each layer, $ l(cm)$, to its density, $ \rho (g/cm ^{3} )$. We have 
\begin{equation}
   z (mg/cm ^{2})= (l_{CH} \times \rho_{CH})+ (l_{Cu} \times \rho_{Cu})
 \label{Eq. target total length}
 \end{equation} 
where $ \rho_{CHCl}$= 1.289 $ (g/cm ^{3} )$ and $ \rho_{Cu}$= 8.92 $ (g/cm ^{3} )$. For CHCl layer we have l$ _{CHCl} $ = 0, 25, and 40 $\mu m$. Unfortunately $ l_{Cu}$ is not defined for the thick targets which means we have to do some assumptions for $ l_{Cu}$. According to Sec.\ref{sec.ESTAR database}, the maximum energy of HE is about 52$ ^{+9} _{-5} $ keV as estimated by the ratio analysis. It means the maximum energy of HE should be about 60 keV. By looking to the ESTAR database for the Cu element, 60 keV equals to range R= 9.46 $ mg/cm^{2} $ which corresponds to a penetration depth (L= R/$ \rho_{Cu}$) of L=10$\mu m$ into the Cu layer. Hence, 10$\mu m$ should be a reasonable choice for the thickness of the Cu layer for thick targets. Now we can assume that X-ray Cu K$ \alpha $ photons are emitted on average at half of this distance, i.e. l$ _{Cu} $ = 5 $\mu m$. Substituting in Eq.\ref{Eq. target total length} gives z(mg/cm $ ^{2} $)= 4.45, 7.7, and 9.7 for  l$ _{CHCl} $ = 0, 25, and 40 $\mu m$ respectively. Now we can plot the number of K$ \alpha $ photons vs the total length of the target
as presented in Fig. \ref{fig.HK-fit}. Fitting according to Eq.\ref{Eq. H-K FIT} allows to find the range, R (g/cm $ ^{2} $) and from this we can find the HE temperature using Eq.\ref{Eq. H-K temperature}. 
\newline
For l$ _{Cu} $ = 5 $\mu m$, the analyses show N$ _{0} $= (5.0 $ \pm $ 3.0)$ \times 10^{11} $  and R= 1.7$ \pm 0.6 $  $mg/cm ^{2} $ which corresponds to HE with energy T$  _{HE} $= 27.7$ \pm $ 5 keV.\\
We can also choose different values for Cu layer thickness such as l$ _{Cu} $ = 2.5 and 1 $\mu m$ to see which HE temperature is estimated by the H-K model. For l$ _{Cu} $ = 2.5 $\mu m$ we have N$ _{0} $= (1.4 $ \pm $ 0.5)$ \times 10^{11} $  and R= 2.7$ \pm 0.9 $  $mg/cm ^{2} $ which corresponds to HE with energy T$  _{HE} $= 35.9$ \pm $ 6.3 keV. For l$ _{Cu} $ = 1 $\mu m$ we have N$ _{0} $= (0.6 $ \pm $ 0.1)$ \times 10^{11} $  and R= 4.1$ \pm 1.3 $  $mg/cm ^{2} $ which corresponds to HE with energy T$  _{HE} $= 45.4$ \pm $ 7.6 keV. We have to mention for l$ _{Cu} \leq $  0.5 $\mu m$, H-K model gives very unrealistic high HE temperature. Accordingly, the maximum HE temperature estimated with the H-K model is less than $ \leq $ 45.4$ \pm $ 7.6 keV. \\
HE energy estimated above by the H-K model are according to the values of $ \mu $, $ \beta $, and b parameters for the Carbon element. For the thick Cu targets without CH front layer, it is instructive to see which T$  _{HE} $ is estimated by the H-K model when we use parameters relevant to the Cu element in Eqs.\ref{Eq. H-K FIT} and \ref{Eq. H-K temperature}. In Harrach \& Kidder paper\cite{Har81}, the values of $ \mu $, $ \beta $, and b parameters are defined only for C, Al, and Au elements. To find the value of each parameter, we use a function which fits the best to the values of C, Al, and Au elements and by interpolating we estimate the value of that parameter for the Cu element. The analysis shows the function which fits the best to the data for all three parameters is a power function in the form of ax$ ^{c} $ where a and c should be defined. For $ \mu $, we find a=0.93 and c=-0.1 which gives $ \mu_{Cu} $= 0.668. For b we find a=1.63 and c=0.56 which gives b$_{Cu} $= 10.8$  \times$10$ ^{-6} $. For $ \beta $ we find a=1.2 and c=0.23 which gives $ \beta_{Cu} $= 2.63. \\
For l$ _{Cu} $ = 5 $\mu m$ and considering the values relevant to Cu in H-K model, the analyses show the same initial number of photons, i.e. N$ _{0} $= (5.0 $ \pm $ 3.0)$ \times 10^{11} $, but the range, now, is R= 3.4$ \pm 1.2 $  $mg/cm ^{2} $ which corresponds to HE with temperature T$  _{HE} $= 31.6$ \pm $ 6 keV which is not too different from the prediction using carbon values in H-K model for l$ _{Cu} $ = 5 $\mu m$. \\
The advantage of H-K model to the simple exponential fit used in Sec.\ref{sec.ESTAR database} is that the simple exponential gives only a single HE energy as an average of whole HE population assuming energetic electrons travelling in forward straight trajectories. Instead, the H-K model includes phenomena such as straggling of electron trajectories and the back reflection of a part of the electrons from the tracer layers. Moreover the H-K model describes the propagation of HE with moderate energies in matter which are strongly affected by the collisions and therefore quickly get a quasi isotropic propagation. \\
Overall, by looking to N$ _{0} $ estimated for l$ _{Cu} $= 2.5 and 5 $\mu m$, we can see they are much higher than the experimental initial number of photons which are less than 4$ \times 10^{10} $, Fig. \ref{fig.HK-fit}, while N$ _{0} $ estimated with l$ _{Cu} $= 1 $\mu m$ is close to the experimental data and also to the one estimated with the ESTAR database. Hence, it seems  T$  _{HE} $= 45.4$ \pm $ 7.6 keV represents more precisely the HE temperature.\\
Our measurements are in agreements with the previous experimental results and the results of other groups obtained in experiment with similar intensities at OMEGA laser facility where they found T$ _{HE} \approx $ 30 keV \cite{The12}.

\section{Conversion efficiency}
Finally we estimate the conversion efficiency, $ \eta $, i.e. the total energy in HE, E$  _{HE}$, with respect to the incident laser energy, E$ _{laser} $

\begin{equation}
\eta= \frac{E_{HE}}{E  _{laser}}
\label{Eq. conversion efficiency} 
\end{equation}  

E$  _{HE}$ is equal to N$  _{e0}\times T_{HE}$, where N$  _{e0}$ is the total number of HE initially produced inside the target and T$_{HE}$ is the average energy of HE. The number of electrons, $ N_{e0} $, is related to the number of emitted K$ \alpha $ photons by Eq.\ref{Eq.Nph by HE}, i.e.

\begin{equation}
 N_{e0} \approx    \frac{N_{0}}{\omega_{K} \: \sigma \: n_{Cu} \: l_{Cu}} 
\label{Eq.Nph vs Ncu}
\end{equation}
\newline
Here, for sake of brevity, we assume mono energetic electrons and we perform the calculation of conversion efficiency with reference only to Cu K$ \alpha $ photons. Hence, $ \omega_{K} $ = 0.44, n$ _{Cu} $ = $8\times 10^{22}cm ^{-3} $, and $ \sigma = 2\times 10^{-22} cm ^{2} $. Following H-K model, Eq.\ref{Eq. H-K FIT}, for l$ _{Cu} $ = 1$ \mu m $, we have $ T_{HE}$= 45.4$ \pm 7.6$ keV and N$  _{0}$=(0.6$ \pm 0.1)\times 10^{11} $. Considering the maximum laser energy  E$ _{laser}  \approx  $ 170 J, Eq.\ref{Eq. conversion efficiency} gives $  \eta \approx $ 0.36$ \pm $0.6 \%. \\
This result is in good agreement with the estimation of similar experiment \cite{Gus05}. They are also in fair agreement with the results of detailed Monte Carlo simulations $  \eta \approx $0.14$ \pm $0.1 \% \cite{bat18}.  

\section{Conclusion}  
We performed an experiment to characterize HE produced in laser-plasma interaction at intensities relevant to SI. We used two types of targets during the experiment for HE studies: thin multilayer targets and thick Cu targets.\\
For thick targets we followed two approaches: first we used an exponential fit to infer HE penetration in target and compared this with data from the ESTAR database. Second we used the model developed by Harrach-Kidder. The two approaches give T$ _{HE,exp} $ = 41.5$ \pm 6 $ keV and T$ _{HE,HK} \leq$ 45.4$ \pm 7.6 $ keV respectively.  \\
For the thin multilayer targets,  we estimated the temperature of HE from the ratio of the number of K$ \alpha $ photons of  Ti to the number of K$ \alpha $ photons of Cu. This analysis estimated HE energies $ \approx$ 52 $ ^{+9} _{-5} $ keV. This value compares well to 41.5$ \pm 6 $ keV. In both cases the basic assumption is replacing the HE distribution with monoenergetic electrons. Instead H-K calculations refer to a Maxwellian distribution. \\
In conclusion from our experiment we see that using 3$ \omega $ irradiation of low Z targets at intensities of interest for shock ignition, we generate a HE distribution with $ \eta \leq$1\% and T$ _{HE} \leq$ 45.4 keV. Although this is low, still there is a significant number of electrons with E$ > $  100 keV in a Maxwell distribution with this temperature. The effect of such HE production in the context of shock ignition must be therefore carefully considered in future works.

\section*{Acknowledgments}
This work has been carried out within the framework of the EUROfusion Enabling Research Project: AWP17-ENR-IFE-CEA-01 \textit{Preparation and Realization of European Shock Ignition Experiments} and has received funding from the Euratom research and training programme 2014-2018 under grant agreement No.633053. The views and opinions expressed herein do not necessarily reflect those of the European Commission.

\nocite{*}


\providecommand{\noopsort}[1]{}\providecommand{\singleletter}[1]{#1}%
\begin{thebibliography}{24}%
\makeatletter
\providecommand \@ifxundefined [1]{%
 \@ifx{#1\undefined}
}%
\providecommand \@ifnum [1]{%
 \ifnum #1\expandafter \@firstoftwo
 \else \expandafter \@secondoftwo
 \fi
}%
\providecommand \@ifx [1]{%
 \ifx #1\expandafter \@firstoftwo
 \else \expandafter \@secondoftwo
 \fi
}%
\providecommand \natexlab [1]{#1}%
\providecommand \enquote  [1]{``#1''}%
\providecommand \bibnamefont  [1]{#1}%
\providecommand \bibfnamefont [1]{#1}%
\providecommand \citenamefont [1]{#1}%
\providecommand \href@noop [0]{\@secondoftwo}%
\providecommand \href [0]{\begingroup \@sanitize@url \@href}%
\providecommand \@href[1]{\@@startlink{#1}\@@href}%
\providecommand \@@href[1]{\endgroup#1\@@endlink}%
\providecommand \@sanitize@url [0]{\catcode `\\12\catcode `\$12\catcode
  `\&12\catcode `\#12\catcode `\^12\catcode `\_12\catcode `\%12\relax}%
\providecommand \@@startlink[1]{}%
\providecommand \@@endlink[0]{}%
\providecommand \url  [0]{\begingroup\@sanitize@url \@url }%
\providecommand \@url [1]{\endgroup\@href {#1}{\urlprefix }}%
\providecommand \urlprefix  [0]{URL }%
\providecommand \Eprint [0]{\href }%
\providecommand \doibase [0]{http://dx.doi.org/}%
\providecommand \selectlanguage [0]{\@gobble}%
\providecommand \bibinfo  [0]{\@secondoftwo}%
\providecommand \bibfield  [0]{\@secondoftwo}%
\providecommand \translation [1]{[#1]}%
\providecommand \BibitemOpen [0]{}%
\providecommand \bibitemStop [0]{}%
\providecommand \bibitemNoStop [0]{.\EOS\space}%
\providecommand \EOS [0]{\spacefactor3000\relax}%
\providecommand \BibitemShut  [1]{\csname bibitem#1\endcsname}%
\let\auto@bib@innerbib\@empty
\bibitem [{\citenamefont {Betti}\ \emph {et~al.}(2007)\citenamefont {Betti},
  \citenamefont {Zhou}, \citenamefont {Anderson}, \citenamefont {Perkins},
  \citenamefont {Theobald},\ and\ \citenamefont {Solodov}}]{Bet07}%
  \BibitemOpen
  \bibfield  {author} {\bibinfo {author} {\bibfnamefont {R.}~\bibnamefont
  {Betti}}, \bibinfo {author} {\bibfnamefont {C.~D.}\ \bibnamefont {Zhou}},
  \bibinfo {author} {\bibfnamefont {K.~S.}\ \bibnamefont {Anderson}}, \bibinfo
  {author} {\bibfnamefont {L.~J.}\ \bibnamefont {Perkins}}, \bibinfo {author}
  {\bibfnamefont {W.}~\bibnamefont {Theobald}}, \ and\ \bibinfo {author}
  {\bibfnamefont {A.~A.}\ \bibnamefont {Solodov}},\ }\bibfield  {title}
  {\enquote {\bibinfo {title} {Shock ignition of thermonuclear fuel with high
  areal density},}\ }\href {\doibase 10.1103/PhysRevLett.98.155001} {\bibfield
  {journal} {\bibinfo  {journal} {Phys. Rev. Lett.}\ }\textbf {\bibinfo
  {volume} {98}},\ \bibinfo {pages} {155001} (\bibinfo {year}
  {2007})}\BibitemShut {NoStop}%
\bibitem [{\citenamefont {Batani}\ \emph
  {et~al.}(2014{\natexlab{a}})\citenamefont {Batani}, \citenamefont {Baton},
  \citenamefont {Casner}, \citenamefont {Depierreux}, \citenamefont
  {Hohenberger}, \citenamefont {Klimo}, \citenamefont {Koenig}, \citenamefont
  {Labaune}, \citenamefont {Ribeyre}, \citenamefont {Rousseaux}, \citenamefont
  {Schurtz}, \citenamefont {Theobald},\ and\ \citenamefont
  {Tikhonchuk}}]{Bat14a}%
  \BibitemOpen
  \bibfield  {author} {\bibinfo {author} {\bibfnamefont {D.}~\bibnamefont
  {Batani}}, \bibinfo {author} {\bibfnamefont {S.}~\bibnamefont {Baton}},
  \bibinfo {author} {\bibfnamefont {A.}~\bibnamefont {Casner}}, \bibinfo
  {author} {\bibfnamefont {S.}~\bibnamefont {Depierreux}}, \bibinfo {author}
  {\bibfnamefont {M.}~\bibnamefont {Hohenberger}}, \bibinfo {author}
  {\bibfnamefont {O.}~\bibnamefont {Klimo}}, \bibinfo {author} {\bibfnamefont
  {M.}~\bibnamefont {Koenig}}, \bibinfo {author} {\bibfnamefont
  {C.}~\bibnamefont {Labaune}}, \bibinfo {author} {\bibfnamefont
  {X.}~\bibnamefont {Ribeyre}}, \bibinfo {author} {\bibfnamefont
  {C.}~\bibnamefont {Rousseaux}}, \bibinfo {author} {\bibfnamefont
  {G.}~\bibnamefont {Schurtz}}, \bibinfo {author} {\bibfnamefont
  {W.}~\bibnamefont {Theobald}}, \ and\ \bibinfo {author} {\bibfnamefont
  {V.}~\bibnamefont {Tikhonchuk}},\ }\bibfield  {title} {\enquote {\bibinfo
  {title} {Physics issues for shock ignition},}\ }\href
  {http://stacks.iop.org/0029-5515/54/i=5/a=054009} {\bibfield  {journal}
  {\bibinfo  {journal} {Nuclear Fusion}\ }\textbf {\bibinfo {volume} {54}},\
  \bibinfo {pages} {054009} (\bibinfo {year} {2014}{\natexlab{a}})}\BibitemShut
  {NoStop}%
\bibitem [{\citenamefont {Perkins}\ \emph {et~al.}(2009)\citenamefont
  {Perkins}, \citenamefont {Betti}, \citenamefont {LaFortune},\ and\
  \citenamefont {Williams}}]{Per09}%
  \BibitemOpen
  \bibfield  {author} {\bibinfo {author} {\bibfnamefont {L.~J.}\ \bibnamefont
  {Perkins}}, \bibinfo {author} {\bibfnamefont {R.}~\bibnamefont {Betti}},
  \bibinfo {author} {\bibfnamefont {K.~N.}\ \bibnamefont {LaFortune}}, \ and\
  \bibinfo {author} {\bibfnamefont {W.~H.}\ \bibnamefont {Williams}},\
  }\bibfield  {title} {\enquote {\bibinfo {title} {Shock ignition: A new
  approach to high gain inertial confinement fusion on the national ignition
  facility},}\ }\href {\doibase 10.1103/PhysRevLett.103.045004} {\bibfield
  {journal} {\bibinfo  {journal} {Phys. Rev. Lett.}\ }\textbf {\bibinfo
  {volume} {103}},\ \bibinfo {pages} {045004} (\bibinfo {year}
  {2009})}\BibitemShut {NoStop}%
\bibitem [{\citenamefont {Ribeyre}\ \emph {et~al.}(2009)\citenamefont
  {Ribeyre}, \citenamefont {Schurtz}, \citenamefont {Lafon}, \citenamefont
  {Galera},\ and\ \citenamefont {Weber}}]{Rib09}%
  \BibitemOpen
  \bibfield  {author} {\bibinfo {author} {\bibfnamefont {X.}~\bibnamefont
  {Ribeyre}}, \bibinfo {author} {\bibfnamefont {G.}~\bibnamefont {Schurtz}},
  \bibinfo {author} {\bibfnamefont {M.}~\bibnamefont {Lafon}}, \bibinfo
  {author} {\bibfnamefont {S.}~\bibnamefont {Galera}}, \ and\ \bibinfo {author}
  {\bibfnamefont {S.}~\bibnamefont {Weber}},\ }\bibfield  {title} {\enquote
  {\bibinfo {title} {Shock ignition: an alternative scheme for hiper},}\ }\href
  {http://stacks.iop.org/0741-3335/51/i=1/a=015013} {\bibfield  {journal}
  {\bibinfo  {journal} {Plasma Physics and Controlled Fusion}\ }\textbf
  {\bibinfo {volume} {51}},\ \bibinfo {pages} {015013} (\bibinfo {year}
  {2009})}\BibitemShut {NoStop}%
\bibitem [{\citenamefont {Koester}\ \emph {et~al.}(2013)\citenamefont
  {Koester}, \citenamefont {Antonelli}, \citenamefont {Atzeni}, \citenamefont
  {Badziak}, \citenamefont {Baffigi}, \citenamefont {Batani}, \citenamefont
  {Cecchetti}, \citenamefont {Chodukowski}, \citenamefont {Consoli},
  \citenamefont {Cristoforetti}, \citenamefont {Angelis}, \citenamefont
  {Folpini}, \citenamefont {Gizzi}, \citenamefont {Kalinowska}, \citenamefont
  {Krousky}, \citenamefont {Kucharik}, \citenamefont {Labate}, \citenamefont
  {Levato}, \citenamefont {Liska}, \citenamefont {Malka}, \citenamefont
  {Maheut}, \citenamefont {Marocchino}, \citenamefont {Nicolai}, \citenamefont
  {O'Dell}, \citenamefont {Parys}, \citenamefont {Pisarczyk}, \citenamefont
  {Raczka}, \citenamefont {Renner}, \citenamefont {Rhee}, \citenamefont
  {Ribeyre}, \citenamefont {Richetta}, \citenamefont {Rosinski}, \citenamefont
  {Ryc}, \citenamefont {Skala}, \citenamefont {Schiavi}, \citenamefont
  {Schurtz}, \citenamefont {Smid}, \citenamefont {Spindloe}, \citenamefont
  {Ullschmied}, \citenamefont {Wolowski},\ and\ \citenamefont {Zaras}}]{Koe13}%
  \BibitemOpen
  \bibfield  {author} {\bibinfo {author} {\bibfnamefont {P.}~\bibnamefont
  {Koester}}, \bibinfo {author} {\bibfnamefont {L.}~\bibnamefont {Antonelli}},
  \bibinfo {author} {\bibfnamefont {S.}~\bibnamefont {Atzeni}}, \bibinfo
  {author} {\bibfnamefont {J.}~\bibnamefont {Badziak}}, \bibinfo {author}
  {\bibfnamefont {F.}~\bibnamefont {Baffigi}}, \bibinfo {author} {\bibfnamefont
  {D.}~\bibnamefont {Batani}}, \bibinfo {author} {\bibfnamefont {C.~A.}\
  \bibnamefont {Cecchetti}}, \bibinfo {author} {\bibfnamefont {T.}~\bibnamefont
  {Chodukowski}}, \bibinfo {author} {\bibfnamefont {F.}~\bibnamefont
  {Consoli}}, \bibinfo {author} {\bibfnamefont {G.}~\bibnamefont
  {Cristoforetti}}, \bibinfo {author} {\bibfnamefont {R.~D.}\ \bibnamefont
  {Angelis}}, \bibinfo {author} {\bibfnamefont {G.}~\bibnamefont {Folpini}},
  \bibinfo {author} {\bibfnamefont {L.~A.}\ \bibnamefont {Gizzi}}, \bibinfo
  {author} {\bibfnamefont {Z.}~\bibnamefont {Kalinowska}}, \bibinfo {author}
  {\bibfnamefont {E.}~\bibnamefont {Krousky}}, \bibinfo {author} {\bibfnamefont
  {M.}~\bibnamefont {Kucharik}}, \bibinfo {author} {\bibfnamefont
  {L.}~\bibnamefont {Labate}}, \bibinfo {author} {\bibfnamefont
  {T.}~\bibnamefont {Levato}}, \bibinfo {author} {\bibfnamefont
  {R.}~\bibnamefont {Liska}}, \bibinfo {author} {\bibfnamefont
  {G.}~\bibnamefont {Malka}}, \bibinfo {author} {\bibfnamefont
  {Y.}~\bibnamefont {Maheut}}, \bibinfo {author} {\bibfnamefont
  {A.}~\bibnamefont {Marocchino}}, \bibinfo {author} {\bibfnamefont
  {P.}~\bibnamefont {Nicolai}}, \bibinfo {author} {\bibfnamefont
  {T.}~\bibnamefont {O'Dell}}, \bibinfo {author} {\bibfnamefont
  {P.}~\bibnamefont {Parys}}, \bibinfo {author} {\bibfnamefont
  {T.}~\bibnamefont {Pisarczyk}}, \bibinfo {author} {\bibfnamefont
  {P.}~\bibnamefont {Raczka}}, \bibinfo {author} {\bibfnamefont
  {O.}~\bibnamefont {Renner}}, \bibinfo {author} {\bibfnamefont {Y.~J.}\
  \bibnamefont {Rhee}}, \bibinfo {author} {\bibfnamefont {X.}~\bibnamefont
  {Ribeyre}}, \bibinfo {author} {\bibfnamefont {M.}~\bibnamefont {Richetta}},
  \bibinfo {author} {\bibfnamefont {M.}~\bibnamefont {Rosinski}}, \bibinfo
  {author} {\bibfnamefont {L.}~\bibnamefont {Ryc}}, \bibinfo {author}
  {\bibfnamefont {J.}~\bibnamefont {Skala}}, \bibinfo {author} {\bibfnamefont
  {A.}~\bibnamefont {Schiavi}}, \bibinfo {author} {\bibfnamefont
  {G.}~\bibnamefont {Schurtz}}, \bibinfo {author} {\bibfnamefont
  {M.}~\bibnamefont {Smid}}, \bibinfo {author} {\bibfnamefont {C.}~\bibnamefont
  {Spindloe}}, \bibinfo {author} {\bibfnamefont {J.}~\bibnamefont
  {Ullschmied}}, \bibinfo {author} {\bibfnamefont {J.}~\bibnamefont
  {Wolowski}}, \ and\ \bibinfo {author} {\bibfnamefont {A.}~\bibnamefont
  {Zaras}},\ }\bibfield  {title} {\enquote {\bibinfo {title} {Recent results
  from experimental studies on laser–plasma coupling in a shock ignition
  relevant regime},}\ }\href {http://stacks.iop.org/0741-3335/55/i=12/a=124045}
  {\bibfield  {journal} {\bibinfo  {journal} {Plasma Physics and Controlled
  Fusion}\ }\textbf {\bibinfo {volume} {55}},\ \bibinfo {pages} {124045}
  (\bibinfo {year} {2013})}\BibitemShut {NoStop}%
\bibitem [{\citenamefont {Batani}\ \emph {et~al.}(2012)\citenamefont {Batani},
  \citenamefont {Malka}, \citenamefont {Schurtz}, \citenamefont {Ribeyre},
  \citenamefont {Lebel}, \citenamefont {Giuffrida}, \citenamefont {Tikhonchuk},
  \citenamefont {Volpe}, \citenamefont {Patria}, \citenamefont {Koester},
  \citenamefont {Labate}, \citenamefont {Gizzi}, \citenamefont {Antonelli},
  \citenamefont {Richetta}, \citenamefont {Nejdl}, \citenamefont {Sawicka},
  \citenamefont {Margarone}, \citenamefont {Krus}, \citenamefont {Krousky},
  \citenamefont {Skala}, \citenamefont {Dudzak}, \citenamefont {Velyhan},
  \citenamefont {Ullshmied}, \citenamefont {Renner}, \citenamefont {Smid},
  \citenamefont {Klimo}, \citenamefont {Atzeni}, \citenamefont {Marocchino},
  \citenamefont {Schiavi}, \citenamefont {Spindloe}, \citenamefont {O'Dell},
  \citenamefont {Vinci}, \citenamefont {Wolowski}, \citenamefont {Badziak},
  \citenamefont {Pysarcizck}, \citenamefont {Rosinski}, \citenamefont
  {Kalinowska},\ and\ \citenamefont {Chodukowski}}]{Bat12}%
  \BibitemOpen
  \bibfield  {author} {\bibinfo {author} {\bibfnamefont {D.}~\bibnamefont
  {Batani}}, \bibinfo {author} {\bibfnamefont {G.}~\bibnamefont {Malka}},
  \bibinfo {author} {\bibfnamefont {G.}~\bibnamefont {Schurtz}}, \bibinfo
  {author} {\bibfnamefont {X.}~\bibnamefont {Ribeyre}}, \bibinfo {author}
  {\bibfnamefont {E.}~\bibnamefont {Lebel}}, \bibinfo {author} {\bibfnamefont
  {L.}~\bibnamefont {Giuffrida}}, \bibinfo {author} {\bibfnamefont
  {V.}~\bibnamefont {Tikhonchuk}}, \bibinfo {author} {\bibfnamefont
  {L.}~\bibnamefont {Volpe}}, \bibinfo {author} {\bibfnamefont
  {A.}~\bibnamefont {Patria}}, \bibinfo {author} {\bibfnamefont
  {P.}~\bibnamefont {Koester}}, \bibinfo {author} {\bibfnamefont
  {L.}~\bibnamefont {Labate}}, \bibinfo {author} {\bibfnamefont {L.~A.}\
  \bibnamefont {Gizzi}}, \bibinfo {author} {\bibfnamefont {L.}~\bibnamefont
  {Antonelli}}, \bibinfo {author} {\bibfnamefont {M.}~\bibnamefont {Richetta}},
  \bibinfo {author} {\bibfnamefont {J.}~\bibnamefont {Nejdl}}, \bibinfo
  {author} {\bibfnamefont {M.}~\bibnamefont {Sawicka}}, \bibinfo {author}
  {\bibfnamefont {D.}~\bibnamefont {Margarone}}, \bibinfo {author}
  {\bibfnamefont {M.}~\bibnamefont {Krus}}, \bibinfo {author} {\bibfnamefont
  {E.}~\bibnamefont {Krousky}}, \bibinfo {author} {\bibfnamefont
  {J.}~\bibnamefont {Skala}}, \bibinfo {author} {\bibfnamefont
  {R.}~\bibnamefont {Dudzak}}, \bibinfo {author} {\bibfnamefont
  {A.}~\bibnamefont {Velyhan}}, \bibinfo {author} {\bibfnamefont
  {J.}~\bibnamefont {Ullshmied}}, \bibinfo {author} {\bibfnamefont
  {O.}~\bibnamefont {Renner}}, \bibinfo {author} {\bibfnamefont
  {M.}~\bibnamefont {Smid}}, \bibinfo {author} {\bibfnamefont {O.}~\bibnamefont
  {Klimo}}, \bibinfo {author} {\bibfnamefont {S.}~\bibnamefont {Atzeni}},
  \bibinfo {author} {\bibfnamefont {A.}~\bibnamefont {Marocchino}}, \bibinfo
  {author} {\bibfnamefont {A.}~\bibnamefont {Schiavi}}, \bibinfo {author}
  {\bibfnamefont {C.}~\bibnamefont {Spindloe}}, \bibinfo {author}
  {\bibfnamefont {T.}~\bibnamefont {O'Dell}}, \bibinfo {author} {\bibfnamefont
  {T.}~\bibnamefont {Vinci}}, \bibinfo {author} {\bibfnamefont
  {J.}~\bibnamefont {Wolowski}}, \bibinfo {author} {\bibfnamefont
  {J.}~\bibnamefont {Badziak}}, \bibinfo {author} {\bibfnamefont
  {T.}~\bibnamefont {Pysarcizck}}, \bibinfo {author} {\bibfnamefont
  {M.}~\bibnamefont {Rosinski}}, \bibinfo {author} {\bibfnamefont
  {Z.}~\bibnamefont {Kalinowska}}, \ and\ \bibinfo {author} {\bibfnamefont
  {T.}~\bibnamefont {Chodukowski}},\ }\bibfield  {title} {\enquote {\bibinfo
  {title} {Preliminary results from recent experiments and future roadmap to
  shock ignition of fusion targets},}\ }\href
  {http://stacks.iop.org/1742-6596/399/i=1/a=012005} {\bibfield  {journal}
  {\bibinfo  {journal} {Journal of Physics: Conference Series}\ }\textbf
  {\bibinfo {volume} {399}},\ \bibinfo {pages} {012005} (\bibinfo {year}
  {2012})}\BibitemShut {NoStop}%
\bibitem [{\citenamefont {Harrach}\ and\ \citenamefont {Kidder}(1981)}]{Har81}%
  \BibitemOpen
  \bibfield  {author} {\bibinfo {author} {\bibfnamefont {R.~J.}\ \bibnamefont
  {Harrach}}\ and\ \bibinfo {author} {\bibfnamefont {R.~E.}\ \bibnamefont
  {Kidder}},\ }\bibfield  {title} {\enquote {\bibinfo {title} {Simple model of
  energy deposition by suprathermal electrons in laser-irradiated targets},}\
  }\href {\doibase 10.1103/PhysRevA.23.887} {\bibfield  {journal} {\bibinfo
  {journal} {Phys. Rev. A}\ }\textbf {\bibinfo {volume} {23}},\ \bibinfo
  {pages} {887--896} (\bibinfo {year} {1981})}\BibitemShut {NoStop}%
\bibitem [{EST(2005)}]{EST05}%
  \BibitemOpen
  \href@noop {} {\enquote {\bibinfo {title} {Estar database},}\ }\bibinfo
  {howpublished} {\url{http://www.nist.gov/pml/data/star/index.cfm}} (\bibinfo
  {year} {2005})\BibitemShut {NoStop}%
\bibitem [{\citenamefont {Batani}\ \emph
  {et~al.}(2014{\natexlab{b}})\citenamefont {Batani}, \citenamefont
  {Antonelli}, \citenamefont {Atzeni}, \citenamefont {Badziak}, \citenamefont
  {Baffigi}, \citenamefont {Chodukowski}, \citenamefont {Consoli},
  \citenamefont {Cristoforetti}, \citenamefont {De~Angelis}, \citenamefont
  {Dudzak}, \citenamefont {Folpini}, \citenamefont {Giuffrida}, \citenamefont
  {Gizzi}, \citenamefont {Kalinowska}, \citenamefont {Koester}, \citenamefont
  {Krousky}, \citenamefont {Krus}, \citenamefont {Labate}, \citenamefont
  {Levato}, \citenamefont {Maheut}, \citenamefont {Malka}, \citenamefont
  {Margarone}, \citenamefont {Marocchino}, \citenamefont {Nejdl}, \citenamefont
  {Nicolai}, \citenamefont {O\&~apos;Dell}, \citenamefont {Pisarczyk},
  \citenamefont {Renner}, \citenamefont {Rhee}, \citenamefont {Ribeyre},
  \citenamefont {Richetta}, \citenamefont {Rosinski}, \citenamefont {Sawicka},
  \citenamefont {Schiavi}, \citenamefont {Skala}, \citenamefont {Smid},
  \citenamefont {Spindloe}, \citenamefont {Ullschmied}, \citenamefont
  {Velyhan},\ and\ \citenamefont {Vinci}}]{Balu14}%
  \BibitemOpen
  \bibfield  {author} {\bibinfo {author} {\bibfnamefont {D.}~\bibnamefont
  {Batani}}, \bibinfo {author} {\bibfnamefont {L.}~\bibnamefont {Antonelli}},
  \bibinfo {author} {\bibfnamefont {S.}~\bibnamefont {Atzeni}}, \bibinfo
  {author} {\bibfnamefont {J.}~\bibnamefont {Badziak}}, \bibinfo {author}
  {\bibfnamefont {F.}~\bibnamefont {Baffigi}}, \bibinfo {author} {\bibfnamefont
  {T.}~\bibnamefont {Chodukowski}}, \bibinfo {author} {\bibfnamefont
  {F.}~\bibnamefont {Consoli}}, \bibinfo {author} {\bibfnamefont
  {G.}~\bibnamefont {Cristoforetti}}, \bibinfo {author} {\bibfnamefont
  {R.}~\bibnamefont {De~Angelis}}, \bibinfo {author} {\bibfnamefont
  {R.}~\bibnamefont {Dudzak}}, \bibinfo {author} {\bibfnamefont
  {G.}~\bibnamefont {Folpini}}, \bibinfo {author} {\bibfnamefont
  {L.}~\bibnamefont {Giuffrida}}, \bibinfo {author} {\bibfnamefont {L.~A.}\
  \bibnamefont {Gizzi}}, \bibinfo {author} {\bibfnamefont {Z.}~\bibnamefont
  {Kalinowska}}, \bibinfo {author} {\bibfnamefont {P.}~\bibnamefont {Koester}},
  \bibinfo {author} {\bibfnamefont {E.}~\bibnamefont {Krousky}}, \bibinfo
  {author} {\bibfnamefont {M.}~\bibnamefont {Krus}}, \bibinfo {author}
  {\bibfnamefont {L.}~\bibnamefont {Labate}}, \bibinfo {author} {\bibfnamefont
  {T.}~\bibnamefont {Levato}}, \bibinfo {author} {\bibfnamefont
  {Y.}~\bibnamefont {Maheut}}, \bibinfo {author} {\bibfnamefont
  {G.}~\bibnamefont {Malka}}, \bibinfo {author} {\bibfnamefont
  {D.}~\bibnamefont {Margarone}}, \bibinfo {author} {\bibfnamefont
  {A.}~\bibnamefont {Marocchino}}, \bibinfo {author} {\bibfnamefont
  {J.}~\bibnamefont {Nejdl}}, \bibinfo {author} {\bibfnamefont
  {P.}~\bibnamefont {Nicolai}}, \bibinfo {author} {\bibfnamefont
  {T.}~\bibnamefont {O\&~apos;Dell}}, \bibinfo {author} {\bibfnamefont
  {T.}~\bibnamefont {Pisarczyk}}, \bibinfo {author} {\bibfnamefont
  {O.}~\bibnamefont {Renner}}, \bibinfo {author} {\bibfnamefont {Y.~J.}\
  \bibnamefont {Rhee}}, \bibinfo {author} {\bibfnamefont {X.}~\bibnamefont
  {Ribeyre}}, \bibinfo {author} {\bibfnamefont {M.}~\bibnamefont {Richetta}},
  \bibinfo {author} {\bibfnamefont {M.}~\bibnamefont {Rosinski}}, \bibinfo
  {author} {\bibfnamefont {M.}~\bibnamefont {Sawicka}}, \bibinfo {author}
  {\bibfnamefont {A.}~\bibnamefont {Schiavi}}, \bibinfo {author} {\bibfnamefont
  {J.}~\bibnamefont {Skala}}, \bibinfo {author} {\bibfnamefont
  {M.}~\bibnamefont {Smid}}, \bibinfo {author} {\bibfnamefont {C.}~\bibnamefont
  {Spindloe}}, \bibinfo {author} {\bibfnamefont {J.}~\bibnamefont
  {Ullschmied}}, \bibinfo {author} {\bibfnamefont {A.}~\bibnamefont {Velyhan}},
  \ and\ \bibinfo {author} {\bibfnamefont {T.}~\bibnamefont {Vinci}},\
  }\bibfield  {title} {\enquote {\bibinfo {title} {Generation of high pressure
  shocks relevant to the shock-ignition intensity regime},}\ }\href {\doibase
  http://dx.doi.org/10.1063/1.4869715} {\bibfield  {journal} {\bibinfo
  {journal} {Physics of Plasmas}\ }\textbf {\bibinfo {volume} {21}},\ \bibinfo
  {eid} {032710} (\bibinfo {year} {2014}{\natexlab{b}}),\
  http://dx.doi.org/10.1063/1.4869715}\BibitemShut {NoStop}%
\bibitem [{\citenamefont {A~Ridgeley}(1989)}]{Rid89}%
  \BibitemOpen
  \bibfield  {author} {\bibinfo {author} {\bibfnamefont {N.-L.~F.}\
  \bibnamefont {A~Ridgeley}},\ }\href {\doibase 10.1117/12.961859} {\enquote
  {\bibinfo {title} {A comparison of kodak def industrex cx and industrex ax
  films at soft x-ray wavelengths},}\ } (\bibinfo {year} {1989})\BibitemShut
  {NoStop}%
\bibitem [{\citenamefont {Renner}\ \emph {et~al.}(2016)\citenamefont {Renner},
  \citenamefont {{\v{S}}m{\'\i}d}, \citenamefont {Batani},\ and\ \citenamefont
  {Antonelli}}]{ren16}%
  \BibitemOpen
  \bibfield  {author} {\bibinfo {author} {\bibfnamefont {O.}~\bibnamefont
  {Renner}}, \bibinfo {author} {\bibfnamefont {M.}~\bibnamefont
  {{\v{S}}m{\'\i}d}}, \bibinfo {author} {\bibfnamefont {D.}~\bibnamefont
  {Batani}}, \ and\ \bibinfo {author} {\bibfnamefont {L.}~\bibnamefont
  {Antonelli}},\ }\bibfield  {title} {\enquote {\bibinfo {title} {Suprathermal
  electron production in laser-irradiated cu targets characterized by combined
  methods of x-ray imaging and spectroscopy},}\ }\href@noop {} {\bibfield
  {journal} {\bibinfo  {journal} {Plasma Physics and Controlled Fusion}\
  }\textbf {\bibinfo {volume} {58}},\ \bibinfo {pages} {075007} (\bibinfo
  {year} {2016})}\BibitemShut {NoStop}%
\bibitem [{\citenamefont {Pisarczyk}\ \emph {et~al.}(2015)\citenamefont
  {Pisarczyk}, \citenamefont {Gus'kov}, \citenamefont {Renner}, \citenamefont
  {Demchenko}, \citenamefont {Kalinowska}, \citenamefont {Chodukowski},
  \citenamefont {Rosinski}, \citenamefont {Parys}, \citenamefont {Smid},
  \citenamefont {Dostal},\ and\ \citenamefont {et~al.}}]{pis15}%
  \BibitemOpen
  \bibfield  {author} {\bibinfo {author} {\bibfnamefont {T.}~\bibnamefont
  {Pisarczyk}}, \bibinfo {author} {\bibfnamefont {S.}~\bibnamefont {Gus'kov}},
  \bibinfo {author} {\bibfnamefont {O.}~\bibnamefont {Renner}}, \bibinfo
  {author} {\bibfnamefont {N.}~\bibnamefont {Demchenko}}, \bibinfo {author}
  {\bibfnamefont {Z.}~\bibnamefont {Kalinowska}}, \bibinfo {author}
  {\bibfnamefont {T.}~\bibnamefont {Chodukowski}}, \bibinfo {author}
  {\bibfnamefont {M.}~\bibnamefont {Rosinski}}, \bibinfo {author}
  {\bibfnamefont {P.}~\bibnamefont {Parys}}, \bibinfo {author} {\bibfnamefont
  {M.}~\bibnamefont {Smid}}, \bibinfo {author} {\bibfnamefont {J.}~\bibnamefont
  {Dostal}}, \ and\ \bibinfo {author} {\bibnamefont {et~al.}},\ }\bibfield
  {title} {\enquote {\bibinfo {title} {Pre-plasma effect on laser beam energy
  transfer to a dense target under conditions relevant to shock ignition},}\
  }\href {\doibase 10.1017/S0263034615000233} {\bibfield  {journal} {\bibinfo
  {journal} {Laser and Particle Beams}\ }\textbf {\bibinfo {volume} {33}},\
  \bibinfo {pages} {221–236} (\bibinfo {year} {2015})}\BibitemShut {NoStop}%
\bibitem [{\citenamefont {{Morace}}\ and\ \citenamefont
  {{Batani}}(2010)}]{Mor10}%
  \BibitemOpen
  \bibfield  {author} {\bibinfo {author} {\bibfnamefont {A.}~\bibnamefont
  {{Morace}}}\ and\ \bibinfo {author} {\bibfnamefont {D.}~\bibnamefont
  {{Batani}}},\ }\bibfield  {title} {\enquote {\bibinfo {title} {{Spherically
  bent crystal for X-ray imaging of laser produced plasmas}},}\ }\href
  {\doibase 10.1016/j.nima.2010.02.130} {\bibfield  {journal} {\bibinfo
  {journal} {Nuclear Instruments and Methods in Physics Research A}\ }\textbf
  {\bibinfo {volume} {623}},\ \bibinfo {pages} {797--800} (\bibinfo {year}
  {2010})}\BibitemShut {NoStop}%
\bibitem [{\citenamefont {Landau}\ and\ \citenamefont
  {Lifshitz}(2013)}]{landa13}%
  \BibitemOpen
  \bibfield  {author} {\bibinfo {author} {\bibfnamefont {L.~D.}\ \bibnamefont
  {Landau}}\ and\ \bibinfo {author} {\bibfnamefont {E.~M.}\ \bibnamefont
  {Lifshitz}},\ }\href@noop {} {\emph {\bibinfo {title} {Quantum mechanics:
  non-relativistic theory}}},\ Vol.~\bibinfo {volume} {3}\ (\bibinfo
  {publisher} {Elsevier},\ \bibinfo {year} {2013})\ Chap.\ \bibinfo {chapter}
  {XVIII}\BibitemShut {NoStop}%
\bibitem [{\citenamefont {Caporaso}\ and\ \citenamefont
  {Wilson}(1979)}]{Cap79}%
  \BibitemOpen
  \bibfield  {author} {\bibinfo {author} {\bibfnamefont {G.~J.}\ \bibnamefont
  {Caporaso}}\ and\ \bibinfo {author} {\bibfnamefont {S.~S.}\ \bibnamefont
  {Wilson}},\ }\href@noop {} {\bibfield  {journal} {\bibinfo  {journal}
  {Lawrence Livermore National Laboratory Report No. UCRL 83308 (unpublished)}\
  } (\bibinfo {year} {1979})}\BibitemShut {NoStop}%
\bibitem [{\citenamefont {Lee}\ and\ \citenamefont {Trainor}(1980)}]{Lee80}%
  \BibitemOpen
  \bibfield  {author} {\bibinfo {author} {\bibfnamefont {Y.}~\bibnamefont
  {Lee}}\ and\ \bibinfo {author} {\bibfnamefont {R.~T.}\ \bibnamefont
  {Trainor}},\ }\href@noop {} {\bibfield  {journal} {\bibinfo  {journal}
  {Lawrence Livermore National Laboratory Report No. UCID-18574-79-4 pp.34-41
  (unpublished)}\ } (\bibinfo {year} {1980})}\BibitemShut {NoStop}%
\bibitem [{\citenamefont {Spencer}(1959)}]{Spe59}%
  \BibitemOpen
  \bibfield  {author} {\bibinfo {author} {\bibfnamefont {L.}~\bibnamefont
  {Spencer}},\ }\href@noop {} {\bibfield  {journal} {\bibinfo  {journal}
  {National Bureau of Standards Monograph 1 (unpublished)}\ } (\bibinfo {year}
  {1959})}\BibitemShut {NoStop}%
\bibitem [{\citenamefont {Batani}\ \emph {et~al.}(2000)\citenamefont {Batani},
  \citenamefont {Davies}, \citenamefont {Bernardinello}, \citenamefont
  {Pisani}, \citenamefont {Koenig}, \citenamefont {Hall}, \citenamefont
  {Ellwi}, \citenamefont {Norreys}, \citenamefont {Rose}, \citenamefont
  {Djaoui},\ and\ \citenamefont {Neely}}]{Bat00}%
  \BibitemOpen
  \bibfield  {author} {\bibinfo {author} {\bibfnamefont {D.}~\bibnamefont
  {Batani}}, \bibinfo {author} {\bibfnamefont {J.~R.}\ \bibnamefont {Davies}},
  \bibinfo {author} {\bibfnamefont {A.}~\bibnamefont {Bernardinello}}, \bibinfo
  {author} {\bibfnamefont {F.}~\bibnamefont {Pisani}}, \bibinfo {author}
  {\bibfnamefont {M.}~\bibnamefont {Koenig}}, \bibinfo {author} {\bibfnamefont
  {T.~A.}\ \bibnamefont {Hall}}, \bibinfo {author} {\bibfnamefont
  {S.}~\bibnamefont {Ellwi}}, \bibinfo {author} {\bibfnamefont
  {P.}~\bibnamefont {Norreys}}, \bibinfo {author} {\bibfnamefont
  {S.}~\bibnamefont {Rose}}, \bibinfo {author} {\bibfnamefont {A.}~\bibnamefont
  {Djaoui}}, \ and\ \bibinfo {author} {\bibfnamefont {D.}~\bibnamefont
  {Neely}},\ }\bibfield  {title} {\enquote {\bibinfo {title} {Explanations for
  the observed increase in fast electron penetration in laser shock compressed
  materials},}\ }\href {\doibase 10.1103/PhysRevE.61.5725} {\bibfield
  {journal} {\bibinfo  {journal} {Phys. Rev. E}\ }\textbf {\bibinfo {volume}
  {61}},\ \bibinfo {pages} {5725--5733} (\bibinfo {year} {2000})}\BibitemShut
  {NoStop}%
\bibitem [{\citenamefont {Theobald}\ \emph {et~al.}(2012)\citenamefont
  {Theobald}, \citenamefont {Nora}, \citenamefont {Lafon}, \citenamefont
  {Casner}, \citenamefont {Ribeyre}, \citenamefont {Anderson}, \citenamefont
  {Betti}, \citenamefont {Delettrez}, \citenamefont {Frenje}, \citenamefont
  {Glebov}, \citenamefont {Gotchev}, \citenamefont {Hohenberger}, \citenamefont
  {Hu}, \citenamefont {Marshall}, \citenamefont {Meyerhofer}, \citenamefont
  {Sangster}, \citenamefont {Schurtz}, \citenamefont {Seka}, \citenamefont
  {Smalyuk}, \citenamefont {Stoeckl},\ and\ \citenamefont {Yaakobi}}]{The12}%
  \BibitemOpen
  \bibfield  {author} {\bibinfo {author} {\bibfnamefont {W.}~\bibnamefont
  {Theobald}}, \bibinfo {author} {\bibfnamefont {R.}~\bibnamefont {Nora}},
  \bibinfo {author} {\bibfnamefont {M.}~\bibnamefont {Lafon}}, \bibinfo
  {author} {\bibfnamefont {A.}~\bibnamefont {Casner}}, \bibinfo {author}
  {\bibfnamefont {X.}~\bibnamefont {Ribeyre}}, \bibinfo {author} {\bibfnamefont
  {K.~S.}\ \bibnamefont {Anderson}}, \bibinfo {author} {\bibfnamefont
  {R.}~\bibnamefont {Betti}}, \bibinfo {author} {\bibfnamefont {J.~A.}\
  \bibnamefont {Delettrez}}, \bibinfo {author} {\bibfnamefont {J.~A.}\
  \bibnamefont {Frenje}}, \bibinfo {author} {\bibfnamefont {V.~Y.}\
  \bibnamefont {Glebov}}, \bibinfo {author} {\bibfnamefont {O.~V.}\
  \bibnamefont {Gotchev}}, \bibinfo {author} {\bibfnamefont {M.}~\bibnamefont
  {Hohenberger}}, \bibinfo {author} {\bibfnamefont {S.~X.}\ \bibnamefont {Hu}},
  \bibinfo {author} {\bibfnamefont {F.~J.}\ \bibnamefont {Marshall}}, \bibinfo
  {author} {\bibfnamefont {D.~D.}\ \bibnamefont {Meyerhofer}}, \bibinfo
  {author} {\bibfnamefont {T.~C.}\ \bibnamefont {Sangster}}, \bibinfo {author}
  {\bibfnamefont {G.}~\bibnamefont {Schurtz}}, \bibinfo {author} {\bibfnamefont
  {W.}~\bibnamefont {Seka}}, \bibinfo {author} {\bibfnamefont {V.~A.}\
  \bibnamefont {Smalyuk}}, \bibinfo {author} {\bibfnamefont {C.}~\bibnamefont
  {Stoeckl}}, \ and\ \bibinfo {author} {\bibfnamefont {B.}~\bibnamefont
  {Yaakobi}},\ }\bibfield  {title} {\enquote {\bibinfo {title} {Spherical
  shock-ignition experiments with the 40 + 20-beam configuration on omega},}\
  }\href {\doibase http://dx.doi.org/10.1063/1.4763556} {\bibfield  {journal}
  {\bibinfo  {journal} {Physics of Plasmas}\ }\textbf {\bibinfo {volume}
  {19}},\ \bibinfo {eid} {102706} (\bibinfo {year} {2012}),\
  http://dx.doi.org/10.1063/1.4763556}\BibitemShut {NoStop}%
\bibitem [{\citenamefont {Gus'kov}\ \emph {et~al.}(2005)\citenamefont
  {Gus'kov}, \citenamefont {Borodziuk}, \citenamefont {Kalal}, \citenamefont
  {Kasperczuk}, \citenamefont {Kondrashov}, \citenamefont {Limpouch},
  \citenamefont {Pisarczyk}, \citenamefont {Pisarczyk}, \citenamefont
  {Rohlena}, \citenamefont {Skala},\ and\ \citenamefont {Ullschmied}}]{Gus05}%
  \BibitemOpen
  \bibfield  {author} {\bibinfo {author} {\bibfnamefont {S.~Y.}\ \bibnamefont
  {Gus'kov}}, \bibinfo {author} {\bibfnamefont {S.}~\bibnamefont {Borodziuk}},
  \bibinfo {author} {\bibfnamefont {M.}~\bibnamefont {Kalal}}, \bibinfo
  {author} {\bibfnamefont {A.}~\bibnamefont {Kasperczuk}}, \bibinfo {author}
  {\bibfnamefont {V.~N.}\ \bibnamefont {Kondrashov}}, \bibinfo {author}
  {\bibfnamefont {J.}~\bibnamefont {Limpouch}}, \bibinfo {author}
  {\bibfnamefont {P.}~\bibnamefont {Pisarczyk}}, \bibinfo {author}
  {\bibfnamefont {T.}~\bibnamefont {Pisarczyk}}, \bibinfo {author}
  {\bibfnamefont {K.}~\bibnamefont {Rohlena}}, \bibinfo {author} {\bibfnamefont
  {J.}~\bibnamefont {Skala}}, \ and\ \bibinfo {author} {\bibfnamefont
  {J.}~\bibnamefont {Ullschmied}},\ }\bibfield  {title} {\enquote {\bibinfo
  {title} {Investigation of shock wave loading and crater creation by means of
  single and double targets in the pals-laser experiment},}\ }\href {\doibase
  10.1007/s10946-005-0016-2} {\bibfield  {journal} {\bibinfo  {journal}
  {Journal of Russian Laser Research}\ }\textbf {\bibinfo {volume} {26}},\
  \bibinfo {pages} {228--244} (\bibinfo {year} {2005})}\BibitemShut {NoStop}%
\bibitem [{\citenamefont {Batani~D.}(2018)}]{bat18}%
  \BibitemOpen
  \bibfield  {author} {\bibinfo {author} {\bibfnamefont {e.~a.}\ \bibnamefont
  {Batani~D.}, \bibfnamefont {Antonelli~L.}},\ }\bibfield  {title} {\enquote
  {\bibinfo {title} {Progress in understanding the role of hot electrons for
  the shock ignition approach to inertial confinement fusion},}\ }\href@noop {}
  {\bibfield  {journal} {\bibinfo  {journal} {Nuclear Fusion, submitted}\ }
  (\bibinfo {year} {2018})}\BibitemShut {NoStop}%
\bibitem [{\citenamefont {{\v{S}}m{\'\i}d}\ \emph {et~al.}(2014)\citenamefont
  {{\v{S}}m{\'\i}d}, \citenamefont {Renner}, \citenamefont {Rosmej},\ and\
  \citenamefont {Khaghani}}]{smid14}%
  \BibitemOpen
  \bibfield  {author} {\bibinfo {author} {\bibfnamefont {M.}~\bibnamefont
  {{\v{S}}m{\'\i}d}}, \bibinfo {author} {\bibfnamefont {O.}~\bibnamefont
  {Renner}}, \bibinfo {author} {\bibfnamefont {F.~B.}\ \bibnamefont {Rosmej}},
  \ and\ \bibinfo {author} {\bibfnamefont {D.}~\bibnamefont {Khaghani}},\
  }\bibfield  {title} {\enquote {\bibinfo {title} {Investigation of x-ray
  emission induced by hot electrons in dense cu plasmas},}\ }\href@noop {}
  {\bibfield  {journal} {\bibinfo  {journal} {Physica Scripta}\ }\textbf
  {\bibinfo {volume} {2014}},\ \bibinfo {pages} {014020} (\bibinfo {year}
  {2014})}\BibitemShut {NoStop}%
\bibitem [{\citenamefont {Santos}, \citenamefont {Parente},\ and\ \citenamefont
  {Kim}(2003)}]{San03}%
  \BibitemOpen
  \bibfield  {author} {\bibinfo {author} {\bibfnamefont {J.~P.}\ \bibnamefont
  {Santos}}, \bibinfo {author} {\bibfnamefont {F.}~\bibnamefont {Parente}}, \
  and\ \bibinfo {author} {\bibfnamefont {Y.-K.}\ \bibnamefont {Kim}},\
  }\bibfield  {title} {\enquote {\bibinfo {title} {Cross sections for k-shell
  ionization of atoms by electron impact},}\ }\href
  {http://stacks.iop.org/0953-4075/36/i=21/a=002} {\bibfield  {journal}
  {\bibinfo  {journal} {Journal of Physics B: Atomic, Molecular and Optical
  Physics}\ }\textbf {\bibinfo {volume} {36}},\ \bibinfo {pages} {4211}
  (\bibinfo {year} {2003})}\BibitemShut {NoStop}%
\bibitem [{\citenamefont {Batani}\ \emph {et~al.}(2011)\citenamefont {Batani},
  \citenamefont {Koenig}, \citenamefont {Baton}, \citenamefont {Perez},
  \citenamefont {Gizzi}, \citenamefont {Koester}, \citenamefont {Labate},
  \citenamefont {Honrubia}, \citenamefont {Antonelli}, \citenamefont {Morace},
  \citenamefont {Volpe}, \citenamefont {Santos}, \citenamefont {Schurtz},
  \citenamefont {Hulin}, \citenamefont {Ribeyre}, \citenamefont {Fourment},
  \citenamefont {Nicolai}, \citenamefont {Vauzour}, \citenamefont {Gremillet},
  \citenamefont {Nazarov}, \citenamefont {Pasley}, \citenamefont {Richetta},
  \citenamefont {Lancaster}, \citenamefont {Spindloe}, \citenamefont {Tolley},
  \citenamefont {Neely}, \citenamefont {Kozlová}, \citenamefont {Nejdl},
  \citenamefont {Rus}, \citenamefont {Wolowski}, \citenamefont {Badziak},\ and\
  \citenamefont {Dorchies}}]{Bat11}%
  \BibitemOpen
  \bibfield  {author} {\bibinfo {author} {\bibfnamefont {D.}~\bibnamefont
  {Batani}}, \bibinfo {author} {\bibfnamefont {M.}~\bibnamefont {Koenig}},
  \bibinfo {author} {\bibfnamefont {S.}~\bibnamefont {Baton}}, \bibinfo
  {author} {\bibfnamefont {F.}~\bibnamefont {Perez}}, \bibinfo {author}
  {\bibfnamefont {L.~A.}\ \bibnamefont {Gizzi}}, \bibinfo {author}
  {\bibfnamefont {P.}~\bibnamefont {Koester}}, \bibinfo {author} {\bibfnamefont
  {L.}~\bibnamefont {Labate}}, \bibinfo {author} {\bibfnamefont
  {J.}~\bibnamefont {Honrubia}}, \bibinfo {author} {\bibfnamefont
  {L.}~\bibnamefont {Antonelli}}, \bibinfo {author} {\bibfnamefont
  {A.}~\bibnamefont {Morace}}, \bibinfo {author} {\bibfnamefont
  {L.}~\bibnamefont {Volpe}}, \bibinfo {author} {\bibfnamefont
  {J.}~\bibnamefont {Santos}}, \bibinfo {author} {\bibfnamefont
  {G.}~\bibnamefont {Schurtz}}, \bibinfo {author} {\bibfnamefont
  {S.}~\bibnamefont {Hulin}}, \bibinfo {author} {\bibfnamefont
  {X.}~\bibnamefont {Ribeyre}}, \bibinfo {author} {\bibfnamefont
  {C.}~\bibnamefont {Fourment}}, \bibinfo {author} {\bibfnamefont
  {P.}~\bibnamefont {Nicolai}}, \bibinfo {author} {\bibfnamefont
  {B.}~\bibnamefont {Vauzour}}, \bibinfo {author} {\bibfnamefont
  {L.}~\bibnamefont {Gremillet}}, \bibinfo {author} {\bibfnamefont
  {W.}~\bibnamefont {Nazarov}}, \bibinfo {author} {\bibfnamefont
  {J.}~\bibnamefont {Pasley}}, \bibinfo {author} {\bibfnamefont
  {M.}~\bibnamefont {Richetta}}, \bibinfo {author} {\bibfnamefont
  {K.}~\bibnamefont {Lancaster}}, \bibinfo {author} {\bibfnamefont
  {C.}~\bibnamefont {Spindloe}}, \bibinfo {author} {\bibfnamefont
  {M.}~\bibnamefont {Tolley}}, \bibinfo {author} {\bibfnamefont
  {D.}~\bibnamefont {Neely}}, \bibinfo {author} {\bibfnamefont
  {M.}~\bibnamefont {Kozlová}}, \bibinfo {author} {\bibfnamefont
  {J.}~\bibnamefont {Nejdl}}, \bibinfo {author} {\bibfnamefont
  {B.}~\bibnamefont {Rus}}, \bibinfo {author} {\bibfnamefont {J.}~\bibnamefont
  {Wolowski}}, \bibinfo {author} {\bibfnamefont {J.}~\bibnamefont {Badziak}}, \
  and\ \bibinfo {author} {\bibfnamefont {F.}~\bibnamefont {Dorchies}},\
  }\bibfield  {title} {\enquote {\bibinfo {title} {The hiper project for
  inertial confinement fusion and some experimental results on advanced
  ignition schemes},}\ }\href
  {http://stacks.iop.org/0741-3335/53/i=12/a=124041} {\bibfield  {journal}
  {\bibinfo  {journal} {Plasma Physics and Controlled Fusion}\ }\textbf
  {\bibinfo {volume} {53}},\ \bibinfo {pages} {124041} (\bibinfo {year}
  {2011})}\BibitemShut {NoStop}%
\end{thebibliography}%


\providecommand{\noopsort}[1]{}\providecommand{\singleletter}[1]{#1}%

\end{document}